\documentclass[a4paper,11pt]{article}

\usepackage{jcappub} 
\usepackage[export]{adjustbox} 
\usepackage[utf8]{inputenc}
\usepackage{color}
\usepackage{graphicx}
\usepackage{amsmath,amsfonts,amssymb}
\usepackage{acronym}
\usepackage{xspace}
\usepackage{multirow}
\usepackage{tabularx}
\usepackage{orcidlink}
\usepackage{ragged2e}
\usepackage{enumitem}
\usepackage{setspace}
\usepackage{caption}
\usepackage{subcaption}
\usepackage{booktabs}
\usepackage{comment}

\usepackage{listings}
\usepackage[ruled,vlined]{algorithm2e}

\SetCommentSty{mycommfont}

\newcommand{\bea}{\begin{equation}\begin{aligned}} 
\newcommand{\eea}{\end{aligned}\end{equation}}
\newcommand{\be}{\begin{equation}}
\newcommand{\ee}{\end{equation}}

\newcommand{\msun}{M_{\odot}}

\newcommand{\td}{{\rm d}}

\definecolor{rossocorsa}{rgb}{0.83, 0.0, 0.0}
\definecolor{tardisblue}{rgb}{0.0, 0.18, 0.53}
\definecolor{TardisBlueAcceso}{RGB}{0, 80, 200}

\usepackage{lettrine}
\usepackage{Zallman}

\makeatletter
\renewcommand{\@seccntformat}[1]{%
  \ifcsname prefix@#1\endcsname
    \csname prefix@#1\endcsname
  \else
    \csname the#1\endcsname\quad
  \fi}
\title{The impact of non-Gaussianity when searching for Primordial Black Holes with LISA}

\author[a]{A.J.~Iovino\orcidlink{0000-0002-8531-5962},}
\author[*,b]{G. Perna\orcidlink{0000-0002-7364-1904},\note[*]{Corresponding author}}
\author[b]{H.~Veerm\"ae\orcidlink{0000-0003-1845-1355}}

\affiliation[a]{Center for Astrophysics and Space Science (CASS), New York University Abu Dhabi, PO Box 129188, Abu Dhabi, UAE}
\affiliation[b]{Keemilise ja Bioloogilise F\"u\"usika Instituut, R\"avala pst. 10, 10143 Tallinn, Estonia}

\emailAdd{a.iovino@nyu.edu}
\emailAdd{gabriele.perna@phd.unipd.it}
\emailAdd{hardi.veermae@cern.ch}

\abstract{LISA can observe cosmological millihertz (mHz) gravitational wave (GW) backgrounds that may offer a decisive test for asteroid-mass primordial black hole (PBH) dark matter (DM). In standard scenarios, failing to detect a scalar-induced gravitational wave (SIGW) background would exclude the last viable window for PBH DM formed through critical collapse. We show that this conclusion becomes much weaker in the presence of astrophysical foregrounds and strongly non-Gaussian primordial density perturbations, by studying how these phenomena affect the link between SIGWs and PBHs, and reevaluate LISA’s sensitivity to asteroid-mass PBHs. In addition, we analyse the interplay between PBHs and SIGWs to gain further insights into the nature of primordial non-Gaussianity. We find that uncertainties in $f_{\rm NL}$ can induce substantial uncertainties in the PBH abundance, which ultimately limits LISA’s capacity to fully probe the asteroid-mass PBH DM window.
}

\begin{document}
\maketitle
\flushbottom

\section{Introduction}
\label{sec:intro}

\lettrine{G}{ravitational Waves} (GWs) provide a unique window into both the early and late Universe. GW signals, in fact, are expected to be generated by a variety of different astrophysical and cosmological processes (see \cite{Caprini:2018mtu,Regimbau:2011rp, Guzzetti:2016mkm} for some reviews), spanning a much wider frequency range, from the extremely low frequencies probed indirectly by Cosmic Microwave Background (CMB) experiments ($\sim 10^{-18}$ Hz) up to the GHz regime \cite{LISACosmologyWorkingGroup:2022jok,Branchesi:2023mws,Abac:2025saz,Aggarwal:2020olq, Aggarwal:2025noe}. To access this broader spectrum, several detectors have been proposed. Among them, ground-based interferometers like the Einstein Telescope (ET)~\cite{Maggiore:2019uih} and Cosmic Explorer (CE)~\cite{Reitze:2019iox} aim to improve sensitivity down to $\sim0.1$ Hz, while the space-based Laser Interferometer Space Antenna (LISA)~\cite{LISA:2017pwj} will target the milli-Hz band. LISA, which is expected to be operational within a decade, will feature a triangular configuration and will probe frequencies that bridge the gap between ground-based interferometers and those accessible via Pulsar Timing Arrays (PTA) or CMB experiments. Importantly, many cosmological models predict a measurable gravitational wave background in this intermediate frequency band, see e.g.~\cite{Maggiore:2007ulw, Guzzetti:2016mkm, Cai:2017cbj, Caprini:2018mtu}. As a result, any detection or non-detection by LISA will carry fundamental implications for the understanding of early-universe processes and fundamental physics~\cite{LISACosmologyWorkingGroup:2022jok, LISA:2022kgy, LISA:2022yao}. 

Among other sources, scalar-induced gravitational waves (SIGW) \cite{Matarrese:1997ay, Baumann:2007zm, Matarrese:1992rp,Carbone:2004iv, Domenech:2020kqm, Domenech:2021ztg, Ananda:2006af, Adshead:2021hnm, Perna:2024ehx, Bruni:1996im, Acquaviva:2002ud} constitute a particularly interesting cosmological GW background detectable by LISA (see also~\cite{Iovino:2025xkq} for a recent physical explanation of the origin of the signal). As the name suggests, these signals are generated from scalar fluctuations, which act as sources for tensor modes at second order in perturbation theory. As the shape and amplitude of the resulting SIGW spectrum are derived from the scalar power spectrum~\cite{Pi:2020otn}, its detection would be fundamental to unveiling the dynamics of the early-Universe on scales much smaller than those probed by Planck and more recently by the ACT and SPT collaborations~\cite{Planck:2018vyg, ACT:2025fju,SPT-3G:2025bzu}. Various inflationary models~\cite{Garcia-Bellido:1996mdl,Kawasaki:2015ppx, Palma:2020ejf}\footnote{Examples of this are single-field scenarios where the potential shows specific features that lead to an ultra-slow-roll (USR) phase~\cite{Ivanov:1994pa,Kinney:1997ne,Inoue:2001zt,Kinney:2005vj,Martin:2012pe,Motohashi:2017kbs,Ozsoy:2019lyy, Karam:2022nym,Balaji:2022dbi,Allegrini:2024ooy,Allegrini:2025jha} as well as multi-field models~\cite{Garcia-Bellido:1996mdl, Kawasaki:2015ppx, Pi:2017gih, Kallosh:2022vha, Braglia:2022phb, Tada:2023pue,Ferrante:2023bgz,Tada:2023fvd,Wang:2024vfv,Iacconi:2024hmg}, and also mechanisms such as preheating or early matter-dominated eras that can produce a strong enhancement of the GW spectrum without the need for a large enhancement of the scalar power spectrum~\cite{Jedamzik:2010hq,Domenech:2024wao,Papanikolaou:2022chm,Pearce:2023kxp,Ballesteros:2024eee}.} in fact, they predict an amplification of the scalar fluctuations around LISA scales with respect to the nearly-flat power spectrum on CMB scales~\cite{Planck:2018jri}, and this enhancement could possibly generate a large amount of SIGW. In addition, SIGWs are directly sourced by the four-point function of the scalar fluctuation and are sensitive to the possible presence of non-Gaussianity (NG) in the scalar perturbation, as it has been recently analysed by~\cite{Cai:2018dig,Unal:2018yaa,Yuan:2020iwf,Adshead:2021hnm,Perna:2024ehx,Zeng:2025cer}. Primordial NG represents a further powerful probe of the early Universe, being sensitive to the dynamics and interactions in the early Universe. Although standard inflationary scenarios predict mild deviations from Gaussian behaviour, other models could generate substantial NG perturbations~\cite{Komatsu:2010hc,Babich:2004gb,Bartolo:2004if}; thus, constraining primordial NG is fundamental for discriminating between models. Nevertheless, Planck has been able to put tight constraints on the amount of NG, typically parametrized by the quantity $f_{\rm NL}$, on large scales~\cite{Planck:2019kim}, but small scales remain unconstrained. 

Interestingly, the same perturbations that generate SIGW are responsible for the formation of primordial BH (PBH)~\cite{Zeldovich:1967lct,Hawking:1971ei,Carr:1974nx,Carr:1975qj,Chapline:1975ojl}. Different mechanisms have been proposed to possibly produce PBHs, but in the standard formation scenario, which we consider in this work, PBHs are formed as a consequence of the gravitational collapse of large overdensities after they re-enter the horizon. In this scenario, the formation of PBHs is boosted if the scalar power spectrum is enhanced, resulting in a non-negligible PBH abundance~\cite{Ivanov:1994pa,Ivanov:1997ia,Blinnikov:2016bxu}. The mass of the resulting GWs is linked to the wavenumber associated with the perturbation responsible for its formation. Thus, to SIGWs in the mHz frequency band correspond PBHs in the asteroidal-mass window, which is, as of today, fully unbounded~\cite{Bartolo:2018evs,Bartolo:2018rku, Yuan:2021qgz,LISACosmologyWorkingGroup:2023njw,Balaji:2022rsy}. Therefore, detecting or setting upper bounds on SIGWs would have fundamental implications for constraining the PBH abundance. 

Previous works have explored how a possible PTA observation of SIGWs would impact constraints on the PBH abundance (see, e.g.,~\cite{Chen:2019xse,Iovino:2024tyg}). In this work, our aim is to understand the implications of SIGW in the LISA frequency range on the abundance of PBHs in the asteroidal mass range. By analysing the possible detection of SIGW over the whole range, we derive the corresponding bounds on the amplitude of the scalar power spectrum, first assuming Gaussianity of the primordial perturbations and then considering the presence of NG. Thus, we derive the bounds on the PBH abundance by studying the effect of different widths of the primordial power spectrum and the presence of astrophysical foregrounds in the LISA range. After that, we discuss the detectability of primordial NG and the mutual implications between SIGWs and PBH. The paper is organised as follows. In Section~\ref{sec:NGs}, we introduce primordial NG, focusing on the local shape, and explore its impact on the computation of SIGWs and PBH abundance, including a brief discussion of the perturbativity criterion. In Section~\ref{sec:relat}, we describe how to relate the projected constraints on SIGWs to the PBH abundance. We then present our results in Section~\ref{sec:analysis}, where we also discuss the uncertainties in LISA’s capability to constrain primordial NG. We conclude in Section~\ref{sec:Conc}.

\section{Non-Gaussian primordial curvature perturbations}
\label{sec:NGs}

The two key phenomena examined here, PBHs and SIGW, both stem from scalar curvature perturbations. For Gaussian perturbations, their statistics are completely specified by the power spectrum, yielding a clear connection between SIGW and PBHs. Non-Gaussianity, however, can obscure this relationship.

Primordial NG indicates departures from the Gaussian behaviour of the quantum fluctuations during inflation, and it is considered a fundamental observable to shed light on the dynamics of the primordial Universe (see  \cite{Komatsu:2010hc,Babich:2004gb} or \cite{Bartolo:2004if} for a review). 
The usual way to account for local primordial NG is to express the non-Gaussian curvature perturbation $\zeta$ in terms of an auxiliary Gaussian field $\zeta_{\rm G}$ as
\bea
\label{eq:zeta}
    \zeta(\vec{x}) = F(\zeta_{\rm G}(\vec{x}))\,,
\eea
which, if the deviation from Gaussianity is small, can be described by the series expansion
\bea
\label{Eq::local_NG}
    \zeta(\vec{x}) 
    = \zeta_{\rm G}(\vec{x}) 
    + \frac{3}{5}f_{\rm NL} \left(\zeta_{\rm G}^2(\vec{x}) - \langle \zeta_G^2\rangle\right)
    + \dots\,,
\eea
where the term $\langle \zeta_G^2\rangle$ has been introduced in order to let $\langle \zeta \rangle=0$. The amount of NG can thus be typically quantified by $f_{\rm NL}$. Its value depends on the specific inflation model considered. Although the Planck Collaboration~\cite{Planck:2019kim} has placed strong constraints on different types of NG, all of which are still compatible with 0, the error bars are still large. We focus in this work on NG of the local type, which is typical of multi-field inflation models (see, e.g.~\cite{Wands:2010af}); moreover, its imprints on the SIGWs have been widely studied, given the feasibility of the calculations~\cite{Cai:2018dig, Adshead:2021hnm,Perna:2024ehx}.

To determine $\zeta$, one must thus specify $F$ the power spectrum of $\zeta_G$. In our analysis, we we will consider a log-normal power spectrum, characterized by a peak frequency $k_*$ and a width $\Delta$ as
\bea\label{eq:PSlog}
    \mathcal{P}_{\zeta_G}(x)
    =
    \frac{A}{\sqrt{2\pi\Delta^2}}\,{\rm exp}\left[-\frac{\ln^2(k/k_*)}{2\Delta^2}\right]\,.
\eea
One could equivalently fix the power spectrum of $\zeta$ instead of the auxiliary $\zeta_{G}$. However, since both the PBH formation and the perturbative expansion for the SIGW are formulated in terms of the auxiliary $\zeta_G$ field, we find it more convenient to start from $\zeta_G$. As long as NG is weak, the difference between the power spectra $\zeta$ and $\zeta_{G}$ must be small by definition.

With Eq.~\eqref{Eq::local_NG}, we will consider only the leading ($f_{\rm NL}$) term of the local expansion. In principle, higher-order powers of $\zeta_{\rm G}$ and a full series of NG parameters, $g_{\rm NL}$, $h_{\rm NL}$, etc., are present. These terms are expected to give a subdominant contribution, as long as perturbativity holds. Perturbativity will be discussed in detail in Sec.~\eqref{sec:pert}. Nevertheless, even if the overall contribution is small, the higher order terms may contribute non-negligibly to specific parts of the SIGW spectrum~\cite{Perna:2024ehx}. Moreover, as was shown in~\cite{Ferrante:2022mui,Iovino:2024sgs}, adopting the local expansion to approximate the imprints of NG for some inflation models could lead to an underestimate of the imprints due to NG, with an error of many orders of magnitude. It is therefore important to be aware of the limitations imposed by perturbativity when working with the expansion~\eqref{Eq::local_NG}.

\subsection{Scalar-induced gravitational waves}
\label{sec:SIGWs}

Primordial curvature perturbations induce the production of a GW background~\cite{Tomita:1975kj,Matarrese:1993zf,Acquaviva:2002ud,Mollerach:2003nq,Ananda:2006af,Baumann:2007zm} at the second order in perturbation theory (see Ref.~\cite{Iovino:2025xkq} for a recent physical explanation of the origin of SIGW). Omitting anisotropic stress, the Einstein equations imply the following Fourier space equation of motion of GWs
\be\label{eom1}
    h''_{\lambda}({\bf k}, \eta) + 2\mathcal{H}h'_{\lambda}({\bf k}, \eta) + k^2h_{\lambda}({\bf k}, \eta) = 4\mathcal{S}_{\lambda}({\bf k}, \eta)\,, 
\ee
where $\lambda$ is the polarization, $'$ is the derivative with respect to the conformal time $\eta$, and $\mathcal{H} \equiv a H$ the comoving Hubble scale. The lowest order contribution to the source term is
\be\label{ftsource}
    \mathcal{S}_{\lambda}({\bf k}, \eta)
    \!=\! \int\frac{\td^3 q}{(2\pi)^{3}}Q_{\lambda}({\bf k},{\bf q})f(|{\bf k}-{\bf q}|,q,\eta)\zeta_{\bf q}\zeta_{{\bf k}-{\bf q}}\,,
\ee
where $Q_{\lambda}$ is a function projecting the scalar momenta into the polarisation tensor $\varepsilon_{ij}^\lambda$ and $f$ contains the linear evolution of the Newtonian potential and the curvature perturbation $\zeta$. Solving Eq.~\eqref{eom1} via the Green's function method gives the tensor modes in the form of a bilinear functional of the curvature perturbation
\be
\label{Eq:h_sol}
    h_{\lambda}({\bf k},\eta) 
    \!=\! 4\int\frac{\td^3 q}{(2\pi)^{3}} Q_{\lambda}({\bf k},{\bf q})I(|{\bf k}-{\bf q}|,q,\eta)\zeta_{{\bf q}}\zeta_{{\bf k}-{\bf q}} \,,
\ee
where the linear evolution is contained in the integration kernel $I$, whose shape depends on the expansion history of the Universe~\cite{Kohri:2018awv,Domenech:2019quo}. We will assume radiation domination for the rest of this paper. However, modifications to it relevant in the LISA range are, for instance, possible due to brief early matter-dominated periods~\cite{Ferreira:1997hj,Pallis:2005bb,Redmond:2018xty,Cai:2020qpu,Co:2021lkc,Gouttenoire:2021jhk,Chang:2021afa,Cai:2023uhc,Domenech:2024rks} or the softening of the equations of state due to a first-order electroweak phase transition~\cite{Caprini:2024hue,Escriva:2023nzn,Escriva:2024ivo,Lewicki:2024ghw}. The SIGW energy density per logarithmic $k$ at time $\eta$ is\footnote{Note that in our convention the tensor perturbation at second order is defined as $h_{ij}/2$.}
\bea
    \Omega_{\rm GW}(k,\eta) 
    = \frac{1}{48} \left(\frac{k}{\mathcal{H}}\right)^2 \sum_{\lambda = +,\times} \overline{\mathcal{P}_{h,\lambda}(k)}\,,
\eea
where the overline denotes a temporal average and the dimensionless GW power spectrum $\mathcal{P}_h(k)$ at the time $\eta$ is defined through the 2-point function
\be\label{eq:h_2P}
    \langle h_{\lambda_{1}}(\mathbf{k}_1,\eta)h_{\lambda_{2}}(\mathbf{k}_2,\eta) \rangle 
    \equiv (2\pi)^3 \delta^3(\mathbf{k}_1 + \mathbf{k}_2) \delta_{\lambda_1\lambda_2} \frac{2\pi^2}{k^3} \mathcal{P}_{h,\lambda_1}(k_1)\,.
\ee
The present differential GW abundance is obtained as
\be
    \Omega_{\rm GW}(k)
    = \Omega_{\rm rad,0} c_{g}(k) \Omega_{\rm GW}(k,\eta_f)\,,
\ee
where $c_{g}(k) \equiv (g_{*,s}(0)/g_{*,s}(k))^{4/3} (g_{*}(k)/g_{*}(0))$ and  $g_{*}(k)$ and $g_{*,s} (k)$ denote the number of energetic and entropic degrees of freedom at the time the mode $k$ exited the horizon and today ($k \to 0$), respectively. We take the former from~\cite{Borsanyi:2016ksw}, obtaining $c_{g}(k) \approx 0.39$ in the LISA range, and $h^2\Omega_{\rm rad,0}=4.2 \cdot 10^{-5}$~\cite{Planck:2018vyg}.  As is evident from Eq. \eqref{Eq:h_sol}, the present day SIGW spectrum is determined by the 4-point function of curvature perturbations. Indeed, plugging Eq.~\eqref{Eq:h_sol} into Eq.~\eqref{eq:h_2P} gives
\bea
    \langle h_{\lambda_{1}}(\mathbf{k}_1,\eta)h_{\lambda_{2}}(\mathbf{k}_2,\eta) \rangle 
    =& 
    16\int \frac{\td^3 q_1}{(2\pi)^{3}} \frac{\td^3 q_2}{(2\pi)^{3}} 
    Q_{\lambda_1}({\bf k_1},{\bf q}_1)Q_{\lambda_2}({\bf k}_2,{\bf q}_2)
    \\
    &\times
    I(|{\bf k}_1-{\bf q}_1|,q_1,\eta) I(|{\bf k}_2-{\bf q}_2|,q_2,\eta)
    \left\langle\zeta_{{\bf q}_1}\zeta_{{\bf k}_1-{\bf q}_1}\zeta_{{\bf q}_2}\zeta_{{\bf k}_2-{\bf q}_2}\right\rangle\,.
\eea

To study the role of NG, it is convenient to consider the usual cumulant (or connected correlation function) expansion and the fact that all moments of Gaussian fields are determined in terms of their one- and two-point functions when implementing the expansion \eqref{Eq::local_NG}. Thus, we can divide the 4-point function into connected and disconnected components
\be
    \left\langle\zeta_{{\bf k}_1}\zeta_{{\bf k}_2}\zeta_{{\bf k}_3}\zeta_{{\bf k}_4}\right\rangle 
    = \left\langle\zeta_{{\bf k}_1}\zeta_{{\bf k}_2}\zeta_{{\bf k}_3}\zeta_{{\bf k}_4}\right\rangle_{\rm dc}
    +
    \left\langle\zeta_{{\bf k}_1}\zeta_{{\bf k}_2}\zeta_{{\bf k}_3}\zeta_{{\bf k}_4}\right\rangle_{\rm c}\,,
\ee
where, since the one-point function vanishes, the disconnected component is constructed from 2-point functions
\be
    \left\langle\zeta_{{\bf k}_1}\zeta_{{\bf k}_2}\zeta_{{\bf k}_3}\zeta_{{\bf k}_4}\right\rangle_{\rm dc}
    \equiv \left
    \langle\zeta_{{\bf k}_1}\zeta_{{\bf k}_2}\right\rangle \left\langle\zeta_{{\bf k}_3}\zeta_{{\bf k}_4}\right\rangle
    + \left\langle\zeta_{{\bf k}_1}\zeta_{{\bf k}_3}\right\rangle \left\langle\zeta_{{\bf k}_2}\zeta_{{\bf k}_4}\right\rangle
    + \left\langle\zeta_{{\bf k}_1}\zeta_{{\bf k}_4}\right\rangle \left\langle\zeta_{{\bf k}_2}\zeta_{{\bf k}_3}\right\rangle\,
\ee
and the remaining connected component encodes the purely non-Gaussian contribution.

Analogously, the SIGW spectrum can be divided into terms sourced by the connected and disconnected components. The SIGW resulting from the disconnected component is given by
\bea
\label{eq:OmegaGW_dc}
    \overline{\mathcal{P}_{h, \rm dc}}(k)
    =32 \int \frac{d^3 q}{(2 \pi)^3} Q_\lambda^2(\mathbf{k}, \mathbf{q}) I^2(q,|\mathbf{k}-\mathbf{q}|, \eta) P_{\zeta}(q) P_{\zeta}(|\mathbf{k}-\mathbf{q}|) \,,
\eea
where the overline denotes a temporal average and it is the only term that contributes when curvature fluctuations are Gaussian. Nevertheless, it will also receive NG corrections from the non-linear terms in the expansion~\eqref{Eq::local_NG}, which cause the power spectrum $P_{\zeta}(q)$ to deviate from the power spectrum of the auxiliary Gaussian field $P_{\zeta_{G}}(q)$~\cite{Adshead:2021hnm,Perna:2024ehx}. The contribution arising from the manifestly non-Gaussian connected component is 
\bea\label{eq:OmegaGW_c}
    \overline{\mathcal{P}_{h, \rm c}}(k)
    = 
& 16 \int \frac{d^3 q_1}{(2 \pi)^{3 / 2}} \int \frac{d^3 q_2}{(2 \pi)^{3 / 2}} Q_\lambda\left(\mathbf{k}, \mathbf{q}_1\right) I\left(\left|\mathbf{k}-\mathbf{q}_1\right|, q_1, \eta\right) Q_\lambda\left(\mathbf{k}, \mathbf{q}_2\right) \\
& \quad \times I\left(\left|\mathbf{k}-\mathbf{q}_2\right|, q_2, \eta\right) T_{\zeta}\left(\mathbf{q}_1, \mathbf{k}-\mathbf{q}_1,-\mathbf{q}_2, \mathbf{q}_2-\mathbf{k}\right)\,,
\eea
where $T_{\zeta}$ is the primordial connected trispectrum, and its shape depends on the type of primordial NGs considered~\cite{Perna:2024ehx}. It receives its first non-vanishing contribution at the order $f_{\rm NL}^2$.

\subsection{Perturbativity}
\label{sec:pert}

In the following, we will consider only the first non-linear term in the expansion Eq.~\eqref{Eq::local_NG}, and we will stick to perturbative $f_{\rm NL}$. Based on the general structure of the local expansion, this roughly amounts to $\left((3/5)f_{\rm NL}\right)^2 A \ll 1$,
where $A$ quantifies the amplitude of curvature perturbations. The condition above is vague. In particular, $f_{\rm NL}^2 A \approx 1$ is already deep in the non-perturbative regime, as we will show below. Thus, as we aim to keep the analysis as model independent as possible, we will adopt the perturbativity condition
\be\label{eq:PerturbCri1}
    A f_{\rm NL}^2\lesssim 0.1\,
\ee
for the rest of this work. The reasoning below indicates that this condition is quite mild as long as $f_{\rm NL}$ can be treated as a viable expansion parameter in the local expansion~\eqref{Eq::local_NG}. We further stress that the condition for perturbativity will vary depending on the model and should thus be reevaluated once a concrete model is specified.

\begin{figure}
    \centering
    \includegraphics[width=0.99\linewidth]{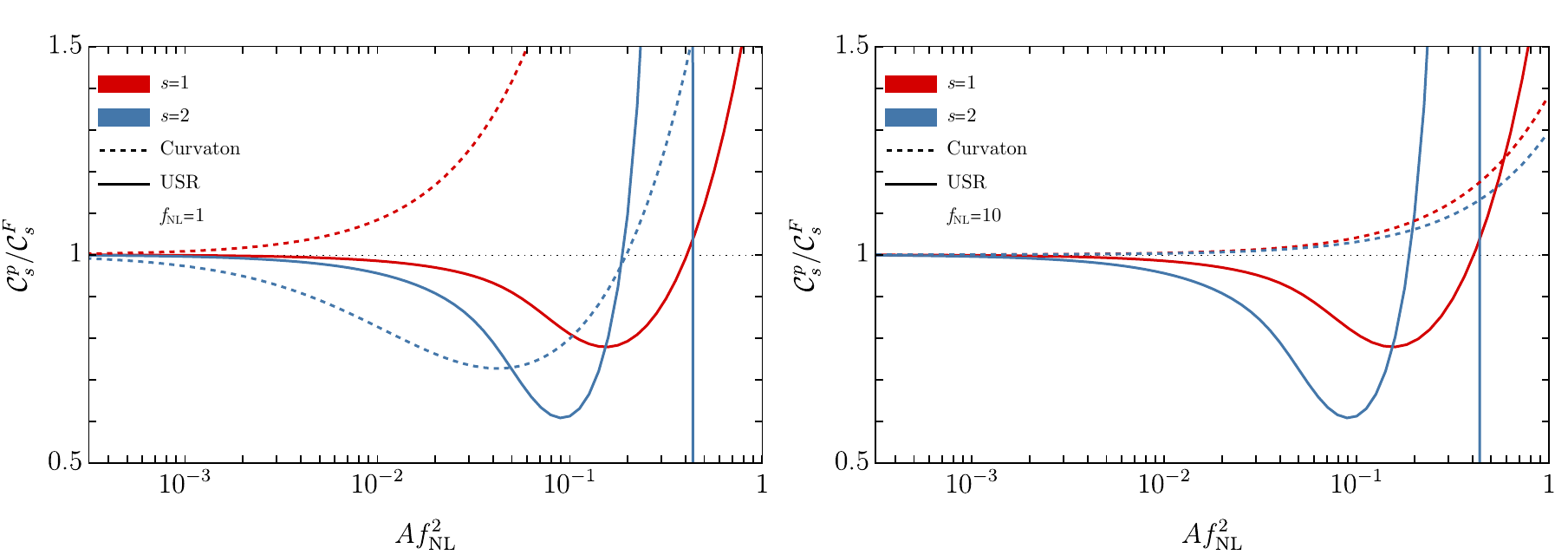}
    \caption{
Ratio of the coefficients $\mathcal{C}_s$ obtained with the perturbative expansion of the non-Gaussian field $\zeta$ as in Eq.~\eqref{Eq::local_NG} and the full functional form $\zeta = F(\zeta_G)$ for two benchmark scenarios: USR (solid lines) and curvaton (dashed lines) models. The colors, red and blue, indicate the order of the coefficient chosen, respectively $s=1$ and $s=2$. The coefficients are shown for $f_{\mathrm{NL}} = 1$ (left) and $f_{\mathrm{NL}} = 10$ (right).}
    \label{fig:Cs}
\end{figure}

Our aim is to provide a back-of-the-envelope estimate of when the usual perturbativity criterion fails in cases in which we know the explicit functional form $F(\zeta_{\rm G})$ as in Eq.~\eqref{eq:zeta}. Looking at the connected four-point function, following Ref.~\cite{Veermae:2026yzz}, the leading order contribution is given by\footnote{In the following, we report only the leading terms of the expansion. At the same perturbative order, an additional contribution proportional to $g_{\rm NL}$ would be present~\cite{Yuan:2023ofl,Li:2023xtl,Perna:2024ehx}; however, we neglect it for simplicity, since we focus exclusively on the $f_{\rm NL}$ contributions.
}

\bea
    \left\langle\zeta_{{\bf k}_1}\zeta_{{\bf k}_2}\zeta_{{\bf k}_3}\zeta_{{\bf k}_4}\right\rangle_{\rm c}
    &= (2\pi)^3 \delta\left( \sum^4_{n = 1}{\bf k}_n \right) (2!)^2\mathcal{C}_1^2\mathcal{C}_2^2  \sum_{\sigma} P_{\sigma_1}P_{\sigma_2}P_{\sigma_3}
    + \dots
    \,.
\eea
We use a shorthand notation for the 
the power spectrum $P_{i} = (2\pi^2/k_i^3)\mathcal{P}_{G}(k_{i})$, and the sum is evaluated over all non-trivial permutations $\sigma$ of the indices. 
The coefficients
\be\label{eq:Cs}
    \mathcal{C}_s 
    = \frac{1}{s! (2A)^{\frac{s+1}{2}}}\int \frac{\td \zeta_G}{\sqrt{\pi}} e^{-\zeta_G^2/(2A)}\, H_s\left(\frac{\zeta_G}{\sqrt{2A}}\right) F(\zeta_G) \,,
\ee
effectively perform a resummation over Wick contractions at a single point, $H_n$ are Hermite polynomials and 
\be
    A \equiv \xi_G(0) 
    =  \int^{\infty}_0 \frac{\td k}{k}\, \mathcal{P}_{G}(k)
\ee
is the variance of fluctuations at a point in space. This non-perturbative approach relies on the conservation of probability when mapping $\zeta$ to $\zeta_G$ and the decomposition of the joint Gaussian probability distribution into Hermite polynomials which allows one to write any $n$-point correlator of $\zeta$ entirely in terms of the coefficients $\mathcal{C}_s$ and combinations of the two-point function of the Gaussian field (for details, see~\cite{Veermae:2026yzz}). Importantly, as \eqref{eq:Cs} does not rely on the series expansion of $F(\zeta_G)$, it can be applied even when $F(\zeta_G)$ is not analytic.

The breakdown of perturbativity can then be estimated by comparing~\eqref{eq:Cs} with the expansion~\eqref{eq:Cs_expansion} for given scenarios. From this, one can quantify the bounds on model parameters, or $f_{\rm NL}$, for which perturbativity fails. To consider perturbativity in detail, let us take the local expansion to an arbitrary order
\bea
\label{Eq::local_NG2}
    \zeta(\vec{x})
    = \sum^{\infty}_{n = 1} F_{{\rm NL},n}  (\zeta_G(\vec{x})^n - \langle\zeta_G(\vec{x})^n\rangle )\,,
\eea
with $F_{\rm NL,1} = 1$ and $F_{\rm NL,2} = (3/5)f_{\rm NL}$, etc. The coefficients \eqref{eq:Cs} can then be expressed by the series by evaluating the integrals (see~\cite{Li:2025met} for an alternative derivation)
\be
    \mathcal{C}_s 
    = \sum_{n\geq 0} \frac{(s+2n)!}{s!(2n)!!} A^n F_{{\rm NL},s+2n}\,. 
\ee
We'd like to know when we can use only the leading term in the expansion Eq.~\eqref{Eq::local_NG2} up to the order of $f_{\rm NL}$ with the only non-vanishing coefficients given by
\be
\label{eq:Cs_expansion}
    \mathcal{C}_1 = 1 + \ldots \quad \textrm{and} \quad
    \mathcal{C}_2 = \frac{3}{5}f_{\rm NL} + \ldots.
\ee

For concreteness, we will consider inflection point inflation featuring an USR phase~\cite{Atal:2018neu, Biagetti:2018pjj, Karam:2022nym} and curvaton models\,\cite{Sasaki:2006kq,Pi:2022ysn} for which the function $\zeta = F(\zeta_G)$ in Eq.~\eqref{eq:zeta} is known. In the USR case~\cite{Atal:2018neu,Tomberg:2023kli},
\bea\label{eq:F_USR}
    \zeta({\bf x}) 
    &= - \frac{1}{\beta}\ln\left(1-\beta \zeta_G({\bf x})\right) 
    = \sum^{\infty}_{n=1} \frac{\beta^{n-1}}{n} \zeta_G({\bf x})^n \quad \mbox{when} \quad|\zeta_G({\bf x})| < \frac{1}{\beta}\,,
\eea
where $\beta$ is related to the subsequent constant-roll phase determined by the inflationary dynamics\footnote{For simplicity, in USR scenarios we neglect the contribution to the PBHs abundance coming from the type II subdominant channel\,\cite{Escriva:2023uko,Wang:2024nmd,Escriva:2025ftp}.}.
For the curvaton, the function $F$ is significantly more complicated~(see, e.g., section 2.4 of \cite{Ferrante:2023bgz}). Nevertheless, the non-Gaussianity is controlled by the curvaton decay parameter $r_{\rm dec}$ also in that case. To relate these examples with the perturbative Eq.~\eqref{Eq::local_NG}, we will match the quadratic terms, which gives, e.g, that $f_{\rm NL} = (5/6)\beta$ in the USR.

In the USR, the expansion of the logarithm in Eq.~\eqref{eq:F_USR} gives the resummed coefficients
\be\label{eq:Cs_USR}
    \mathcal{C}_s 
    = \sum_{n\geq 0} \frac{(s+2n-1)!}{s!(2n)!!} A^n \beta^{s+2n-1}\,.
\ee
The radius of convergence of this series is 0, i.e., it diverges. Thus, to obtain a comparison with the non-perturbative expression, we must compute the integral in Eq.~\eqref{eq:Cs} numerically. Nevertheless, the appearance of divergent perturbative series is commonplace in field theory and~\eqref{eq:Cs_USR} should be interpreted as an asymptotic series whose breakdown can be estimated by when the terms begin to grow. By using the Stirling approximation, we find that terms in the series~\eqref{eq:Cs_USR} begin to grow when $n \gtrsim 1/(2 A\beta^{2})e^{-(s-1)/n}$, thus demanding that at least the first 3 terms in the series decrease at the order $f_{\rm NL}$ ($s=2$), we find that $A\beta^2 \lesssim 0.12$, which translates to
\be
    A f_{\rm NL}^2 \lesssim 0.08\,.
\ee
A similar conclusion was reached in Ref.~\cite{Iovino:2024sgs}.

A numerical comparison is performed in Fig.~\ref{fig:Cs}, where we show the ratio between the coefficients computed using the power series ansatz\footnote{In the numerical integrals, we follow~\cite{Iovino:2024sgs} and use $\zeta({\bf x}) = - \frac{1}{\beta}\ln\left|1-\beta \zeta_G({\bf x})\right|$ when $\zeta_G({\bf x}) \geq 1/\beta$.}, Eq.~\eqref{eq:Cs_expansion}, and the expansion, Eq.~\eqref{eq:Cs}, arising from the curvaton and USR models for $f_{\rm NL} = 1$ and $f_{\rm NL} = 10$.

The estimate of perturbativity is somewhat arbitrary. We choose a criterion that quantifies how accurately the perturbative estimate $\mathcal{C}_s^{ p}$ in Eq.~\eqref{eq:Cs_expansion} matches the non-perturbative resummed estimate $\mathcal{C}_s^{F}$ in Eq.~\eqref{eq:Cs}. For concreteness, we characterize perturbativity by demanding that
\be
    2/3 < \mathcal{C}_s^{p} / \mathcal{C}_s^{F} < 3/2,
\ee
and the breakdown of perturbativity is indicated by the violation of this condition. The perturbativity bounds are shown in Fig.~\ref{fig:Abu}.  We find that, for the USR case, it breaks down for
$A f^2_{\rm NL} \gtrsim 0.06$, which is in good agreement with Eq.~\eqref{eq:PerturbCri1},
while for the curvaton case, it depends on $f_{\rm NL}$.

Indeed, for $f_{\rm NL}=1$, the curvaton model becomes non-perturbative at values comparable to USR, namely when $A f_{\rm NL}^2 \gtrsim 0.06$. For larger $f_{\rm NL} \gg 1$, this bound is much weaker, requiring only $A f_{\rm NL}^2 \gtrsim 1$. Curiously, we also find that the perturbativity criterion becomes much stronger when $|f_{\rm NL}| \ll 1$. Although this result seems counterintuitive, it is possible to understand it by considering the intricate shape of $F$ for the curvaton. Importantly, the curvaton does not possess a Gaussian limit, and the cubic and higher order terms do not vanish when $f_{\rm NL} \to 0$. This can already be observed when considering the next-to-leading order, that is, $g_{\rm NL}$. In terms of the curvaton decay parameter $r_{\rm dec}$, we have that~\cite{Sasaki:2006kq}
\be
    f_{\rm NL} = \frac{5}{3}\left(\frac{3}{4r_{\rm dec}} - 1 - \frac{r_{\rm dec}}{2}\right),
    \qquad
    g_{\rm NL} = \frac{25}{54} \left(-\frac{9}{r_{\rm dec}} +\frac{1}{2} +10 r_{\rm dec} +3 r_{\rm dec}^2\right)\,,
\ee
where $r_{\rm dec} \in [0,1]$. Expressing $g_{\rm NL}$ in terms of $f_{\rm NL}$ then gives the following asymptotics
\be
    \mbox{curvaton NLO:} \qquad
    g_{\rm NL} \sim 
    \left\{\begin{array}{lc}
    2 f_{\rm NL}^2\,, & \qquad \,\,f_{\rm NL} \ll -1\,,
    \\
    -3.8-4.1 f_{\rm NL}\,, & \qquad |f_{\rm NL}| \ll 1\,,\quad
    \\
    - f_{\rm NL}/3\,, & \qquad f_{\rm NL} \gg 1 \,.
    \end{array}\right.
\ee
Thus, for $f_{\rm NL} < 0$, the $g_{\rm NL}$ contribution increases sharply, and perturbativity is quickly broken. Indeed, $g_{\rm NL}$ is already sizeable near $f_{\rm NL} \approx 0$, signalling a loss of perturbativity even at small $f_{\rm NL}$. Conversely, for $f_{\rm NL} \gg 1$, $g_{\rm NL}$ grows only linearly with $f_{\rm NL}$, so perturbativity can be maintained more easily than in the USR scenario, where $g_{\rm NL} \propto f_{\rm NL}^2$. In conclusion, unlike USR, where higher order NG terms were proportional to powers of $f_{\rm NL}$, the unexpected behaviour of the curvaton stems from the more complex structure of the local expansion \eqref{Eq::local_NG2}.

All in all, although we use the condition \eqref{eq:PerturbCri1}, which we expect to hold when the local expansion is organised neatly into growing orders of the expansion parameter, represented by $f_{\rm NL}\sqrt{A}$, we caution the reader to take the perturbativity condition with a grain of salt. As in the example of curvaton, we saw that it can be strengthened when $f_{\rm NL}$ is small and relaxed even beyond the naive expectation $(3/5)^2f_{\rm NL}^2 A < 0$ when $f_{\rm NL}$ is large and positive.

\subsection{Primordial black hole abundance}
\label{sec:PBHs}

At horizon re-entry, sufficiently large density fluctuations can collapse and form a black hole. The mass of the resulting PBH follows a critical scaling law~\cite{Choptuik:1992jv, Niemeyer:1997mt, Niemeyer:1999ak}
\be
    M_{\rm PBH} = \mathcal{K} M_{\rm H} (\mathcal{C} - \mathcal{C}_{\rm th})^{\gamma}\,,
\ee
where $\mathcal{C}$ is the so called compaction function, generically defined as twice the local mass excess over the areal radius, and
\be \label{eq:M_k}
    M_{\rm H}
    \approx 14 \msun \left[\frac{k}{10^6{\rm Mpc}^{-1}} \right]^{-2}\, \left[\frac{g_{*,s}(k)^4 g_{*}(k)^{-3}}{106.25}\right]^{-1/6} \,
\ee  
is the mass contained within a Hubble horizon corresponding to a comoving scale $k$. We take into account the dependence of the threshold on the power spectrum shape\footnote{The threshold depends also on primordial non-Gaussianities~\cite{Kehagias:2019eil,Escriva:2022pnz}. However, since corrections to the threshold have so far been computed only for monochromatic power spectra—which represent unphysical situations—and are generally of the order of a few percent for positive non-Gaussianities (which are the focus of this work), we neglect this effect for simplicity.}, following Ref.~\cite{Musco:2020jjb}, while, since their tuning is negligible in the subsequent analysis, we fix for simplicity $\mathcal{K}=4.4$ and $\gamma=0.38$. The fraction of energy density $\beta_k(M_{\rm PBH})$ per logarithmic PBH mass  due to PBHs can be estimated as 
\be\label{eq:betak}
    \beta_k(M_{\rm PBH})
    = \int_{\mathcal{C}_{\rm th}} \! \td\mathcal{C} \, P_k(\mathcal{C}) \frac{M_{\rm PBH}}{M_{\rm H}}  \delta\left[ \ln\frac{M_{\rm PBH}}{M_{\rm PBH}(\mathcal{C})} \right]\!,
\ee
where $P_k(\mathcal{C})$ denotes the probability that a BH forms in the Hubble patch. The PBH mass function can be obtained directly from the collapse fraction:
\bea\label{eq:df_PBH}
    \!\frac{\td f_{\rm PBH}}{\td \ln M_{\rm PBH}}
&   \!=\! \frac{1}{\Omega_{\rm DM}}\int \frac{\td M_{\rm H}}{M_{\rm H}} \, \beta_k(M_{\rm PBH} ) \left(\frac{M_{\rm eq}}{M_{\rm H}}\right)^{1/2} \!\!,
\eea
where $M_{\rm eq} \approx 2.8\times 10^{17}\,\,M_{\odot}$ is the horizon mass at the time of matter-radiation equality and $\Omega_{\rm  DM} = 0.12h^{-2}$ is the cold dark matter density~\cite{Planck:2018jri}. To characterise the PBH population, we will consider the PBH abundance and the mean PBH mass,
\bea
   f_{\rm PBH} 
   &= \int \frac{\td M_{\rm PBH}}{M_{\rm PBH}} \frac{\td f_{\rm PBH}}{\td \ln M_{\rm PBH}}\, , \\
   \langle M_{\rm PBH} \rangle 
   &= f_{\rm PBH} \left(\int \frac{\td M_{\rm PBH}}{M_{\rm PBH}^2} \frac{\td f_{\rm PBH}}{\td \ln M_{\rm PBH}}\right)^{-1} \,,
\eea
where the mean PBH mass is computed with respect to the PBH number density (see e.g.~\cite{Andres-Carcasona:2024wqk}). We follow Ref.~\cite{Carr:2017jsz} to recompute the already existing constraints with extended mass functions.

To estimate the mass function, we typically assume a peaked primordial power spectrum. The shape of this peak will non-trivially affect the PBH mass distribution and the PBH abundance. Nevertheless, assuming that expansion is dominated by radiation, the dependence on the position of the peak can be described by a simple scaling law. In detail, given a curvature power spectrum of the form $\mathcal{P}_\zeta(k/k_{*})$, the abundance and masses of PBHs scale as
\bea
    f_{\rm PBH} (k_{*})
    &= f_{\rm PBH}(k_{\rm ref}) \left(\frac{k_{*}}{k_{\rm ref}}\right)\,,
    \\
    \langle M_{\rm PBH} \rangle (k_{*})
    &= \langle M_{\rm PBH} \rangle (k_{\rm ref}) \left(\frac{k_{*}}{k_{\rm ref}}\right)^{-2}\,,
\eea
given that the other parameters of the spectrum are kept fixed.

Common approaches in the literature to determine the PBH abundance are called threshold statistics (TS) and peaks theory (PT). However, the shape and origin of $P_k(\mathcal{C})$ in these two approaches are different (see Ref.~\cite{Iovino:2024tyg} for details).
\begin{figure}
    \centering
    \includegraphics[width=0.99\linewidth]{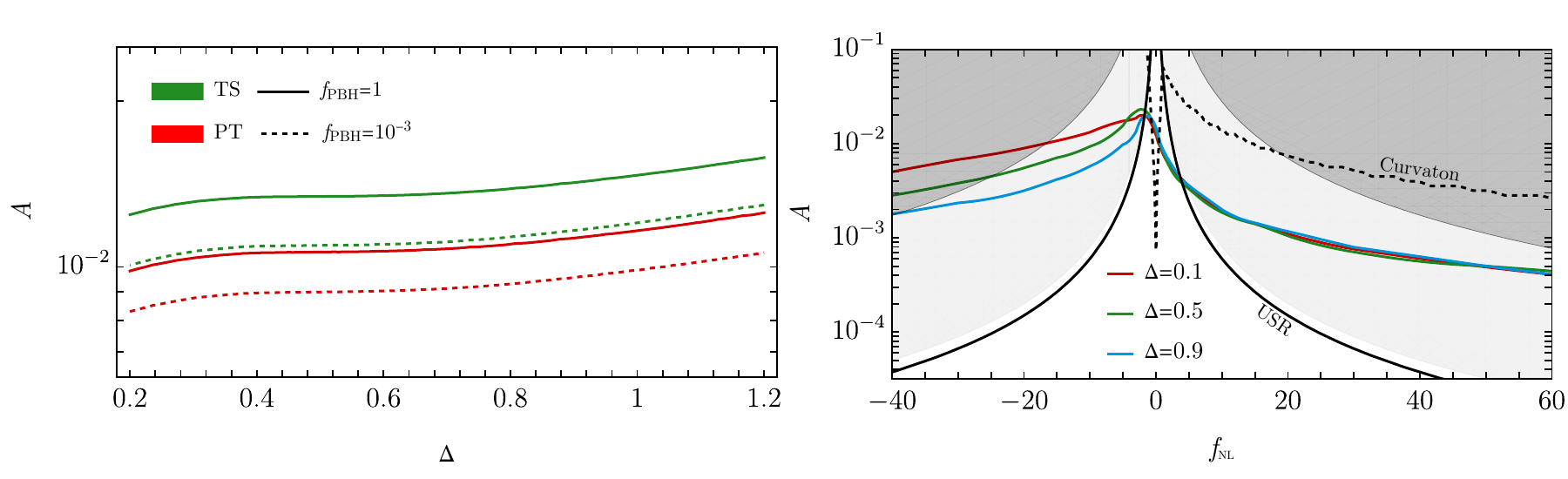}
    \caption{\textbf{ Left panel:}  Amplitude $A$ of the power spectrum, e.g. Eq.~\ref{eq:PSlog}, that is required to get respectively $f_{\rm PBH}=1$ (solid lines) and $f_{\rm PBH}=10^{-3}$ (dashed lines) using TS (red) and the PT (green) as a function of the power spectrum width $\Delta$, assuming negligible primordial NGs, i.e $f_{\rm NL}=0$. \textbf{Right panel:} Amplitude $A$ required to obtain $f_{\rm PBH} = 1$ for different $f_{\rm NL}$ and different benchmark cases for the width $\Delta$, within the TS approach. The excluded regions by the perturbativity condition, Eq.~\eqref{eq:PerturbCri1}, are reported in grey while the perturbativity condition for the USR and curvaton cases are shown as solid and dashed black lines, respectively. In dark grey we report region excluded by the naive condition usually adopted in literature, i.e. $(3/5)^2f_{\rm NL}^2 A < 1$. In the left and right panels, we use the benchmark values $k_*=10^{12.5}$ $\textrm{Mpc}^{-1}$ and $k_*=10^{8}$ $\textrm{Mpc}^{-1}$, respectively.}
    \label{fig:Abu}
\end{figure}
It is well known that the TS approach does not agree with the PT already at the Gaussian level (see, e.g., Refs.~\cite{Green:2004wb, Young:2014ana, DeLuca:2019qsy}).
As shown in the left panel of Fig.~\ref{fig:Abu}  this discrepancy can be absorbed with a small fine tuning of the amplitude and if we fix the amplitude, in the two approaches the PBH abundance differ by a few orders of magnitude $f^{\rm PS}_{\rm PBH}/f^{\rm TS}_{\rm PBH}\lesssim 10^{5}$.
A technical drawback of PT is that there is no clear way to include NGs in the computation of the abundance in the case of a generic functional form for the curvature perturbation field as in Eq.~\eqref{eq:zeta} or for very large NGs. Moreover, as we will see in the next sections, LISA can constrain $f_{\rm PBH}\ll 10^{-40}$. Hence, we limit our analysis to the TS approach, assuming that approximately similar results hold for the PT approach.

As illustrated in the right panel of Fig.~\ref{fig:Abu}, the degeneracy between the amplitude $A$ of the primordial curvature power spectrum and the NG parameter $f_{\mathrm{NL}}$ significantly influences the threshold at which perturbativity breaks down. Being model agnostic, for positive NG, we find that this threshold is largely insensitive to the spectral shape, with the perturbativity condition in Eq.~\eqref{eq:PerturbCri1} being preserved only in the range $f_{\mathrm{NL}}\sim[-2,5]$, regardless of the width $\Delta$ of the power spectrum.

The situation slightly changes when we account for the  perturbativity condition in the USR models. Indeed, in the this type of scenario, since the condition for perturbativity reads $Af^2_{\rm NL}\lesssim0.06$, a similar small range $-2\lesssim f_{\rm NL}\lesssim 5$ is perturbative when $f_{\rm PBH} = 1$, as can be seen in Fig.~\ref{fig:Abu}. In curvaton models, the constraints are more relaxed, but not when $f_{\rm NL}$ is small. This is because, for the curvaton, the condition $f_{\rm NL} = 0$ does not imply that $\zeta$ is Gaussian, as was shown in section~\ref{sec:pert}.

Looking at the amplitude required to achieve $f_{\rm PBH}\simeq1$, we observe that the maximum achievable amplitude $A$ of the power spectrum typically peaks around $f_{\mathrm{NL}} \approx -2$, in agreement with Ref.~\cite{Franciolini:2023pbf,Gouttenoire:2025jxe}, with only a mild sensitivity to the spectral width, especially for positive NG. Notably, the curves in Fig.~\ref{fig:Abu} display a steep slope between the purely Gaussian case and moderately positive NG (up to $f_{\mathrm{NL}} \approx 10$). This implies that even relatively small deviations from Gaussianity can have a significant effect on the resulting PBH abundance. This is due to the exponential sensitivity of PBH formation to the tail of the probability distribution for curvature perturbations. As a result, mild positive NG can amplify the abundance by several orders of magnitude, an effect that is especially pronounced for narrow power spectra, where the contribution from rare, large fluctuations is dominant. This sensitivity underscores the importance of accurately modelling non-Gaussian effects when translating constraints from GW observations into bounds on PBH dark matter; particularly, as we will see in the next sections, in the asteroid-mass window probed by experiments such as LISA. Moreover we stress that adopting the local expansion to approximate the imprints of NG for some inflation models leads to a large error of many orders of magnitude in the PBH abundance estimate\,\cite{Ferrante:2022mui,Ezquiaga:2019ftu,Ezquiaga:2022qpw}. 

\begin{figure}
    \centering
    \includegraphics[width=0.99\linewidth]{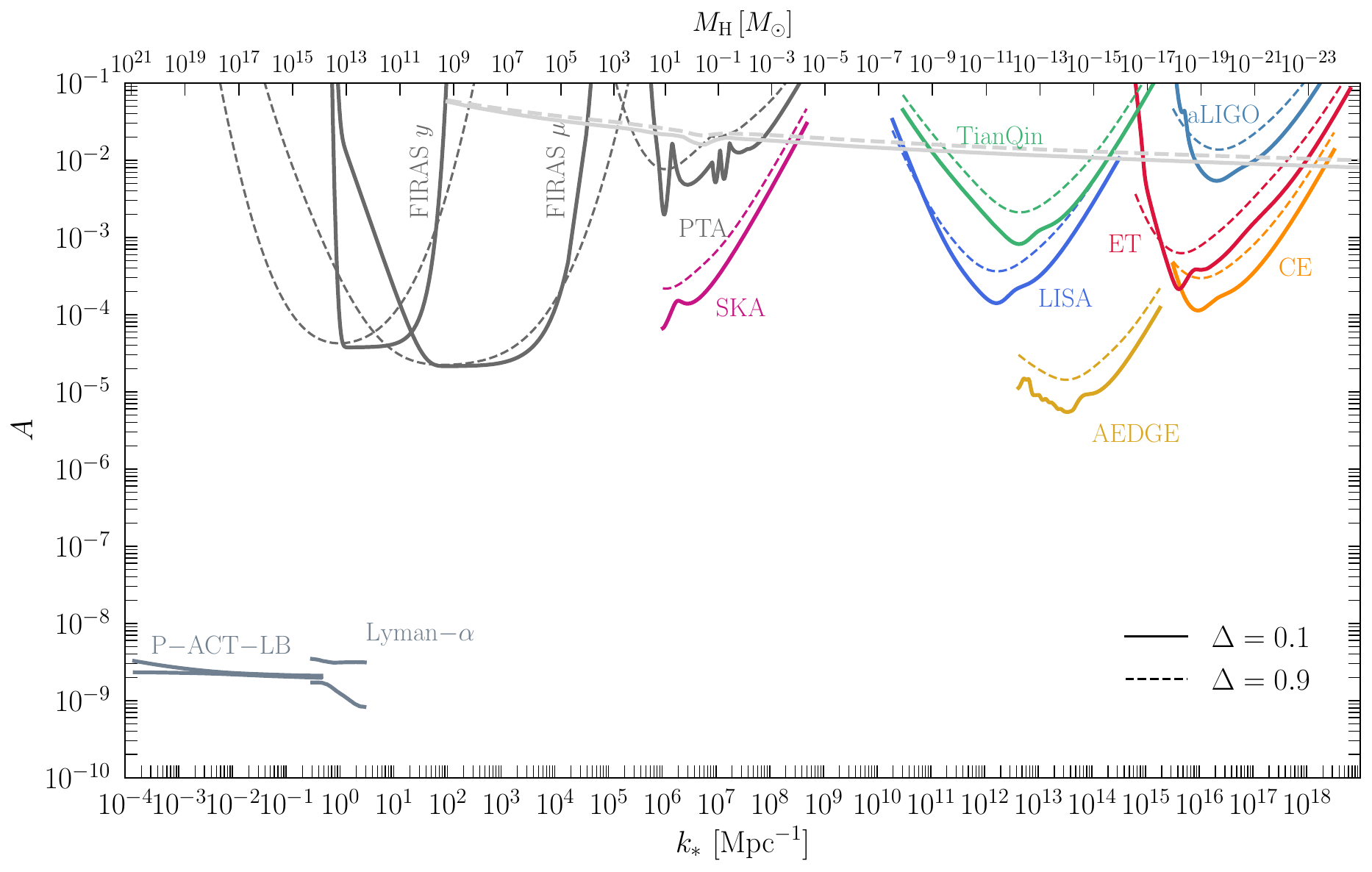}
    \caption{Constraints and prospective sensitives on the primordial scalar power spectrum for current and future GW experiments. The curvature perturbations are assumed Gaussian statistics and to follow log-normal power spectra with for two different widths, $\Delta =0.1$ (solid lines) and $\Delta =0.9$ (dashed lines). The existing constraints for \textbf{Planck+ACT}\,\cite{ACT:2025tim} and \textbf{Ly -$\alpha$}\, \cite{lyman} are taken directly from their respective analyses, while for  \textbf{FIRAS}\,\cite{Chluba:2012we,Chluba:2013dna,Iovino:2024tyg} and \textbf{NANOGrav15}\,\cite{NANOGrav:2023gor,NANOGrav:2023hde,Iovino:2024tyg} they are computed using the latest data and assuming our template for the primordial power spectrum. The sensitivities are shown for \textbf{SKA}~\cite{Janssen:2014dka}, \textbf{LISA}~\cite{LISA:2017pwj}, \textbf{AEDGE}~\cite{AEDGE:2019nxb}, \textbf{TianQin}~\cite{TianQin:2015yph}, \textbf{ET}~\cite{Maggiore:2019uih}, \textbf{CE}~\cite{Reitze:2019iox} and \textbf{aLIGO}~\cite{LIGOScientific:2016wof} estimated using ${\rm SNR}_{\rm th}=10$ and an observation time of $1$ year. The light grey lines indicate the upper bound on $A$ due to the PBH overproduction (assuming the TS formalism).}
    \label{fig:MasterPower}
\end{figure}

\section{Relating cosmological GWs to PBHs}
\label{sec:relat}

\subsection{Constraints on the scalar power spectrum}

As stated above, in our analysis, we will consider the log-normal power spectrum \eqref{eq:PSlog}, characterised by the amplitude $A$, peak frequency $k_*$ and width $\Delta$.
To estimate the prospective sensitivity of future detectors, we use the signal-to-noise-ratio (SNR)
\bea
\label{Eq:SNR}
    {\rm SNR} = \sqrt{T_{\rm obs}} \sqrt{\int df \left( \frac{h^2 \Omega_{\rm GW}(f)}{h^2 \Omega_{\rm noise}(f)}\right)^2}\,,
\eea
where
\bea
    h^2 \Omega_{\rm noise}(f) = \frac{4 \pi^2 f^3}{3 H_0^2/h^2}S_h(f)
\eea
and $S_h(f)$ is the effective sensitivity associated with each detector. A signal is considered detectable when the SNR exceeds the threshold ${\rm SNR}_{\rm thr} = 10$.

Fig.~\ref{fig:MasterPower} shows the prospective sensitivities for Gaussian curvature perturbations assuming an observation time $T_{\rm obs}=1$ yr\footnote{Prospects for a different SNR and $T_{\rm obs}$ can be obtained by multiplying the bounds on $A$ by $({\rm SNR}^{\rm new}/{{\rm SNR}^{\rm old}})^{1/2} (T^{\rm new}_{\rm obs}/T^{\rm old}_{\rm obs})^{-1/4}$.}  
together with current constraints on the amplitude of the primordial power spectrum. At the largest scales, we report the constraints from Planck+ACT~\cite{ACT:2025tim} and Lyman-$\alpha$ (Ly-$\alpha$)~\cite{Bird:2010mp} which do not depend on our templates for the power spectrum. For higher $k$, the constraints and sensitivities are evaluated for two different log-normal curvature power spectra with a narrow ($\Delta=0.1$) and a wide peak ($\Delta=0.9$). The constraints from Firas~\cite{Chluba:2012we,Chluba:2013dna} and NANOGrav~\cite{NANOGrav:2023gor,NANOGrav:2023hde} are obtained as in~\cite{Iovino:2024tyg}. Finally, we show the projected sensitivities of SKA~\cite{Janssen:2014dka}, LISA~\cite{LISA:2017pwj}, AEDGE~\cite{AEDGE:2019nxb}, TIANQin~\cite{TianQin:2015yph}, ET~\cite{Maggiore:2019uih}, {CE}~\cite{Reitze:2019iox} and {aLIGO}~\cite{LIGOScientific:2016wof}. For the prospects from PTA~\cite{Cecchini:2025oks} and SKA~\cite{Babak:2024yhu} we neglected astrophysical foregrounds due to SMBH binaries. 

Finally, the light grey line corresponds to $f_{\rm PBH}=1$ and thus shows the constraint from PBH overproduction assuming PBH formation from Gaussian perturbations and TS. 
On the upper $x$-axis we show the horizon mass Eq.~\eqref{eq:M_k} corresponding to $k_{*}$. It is closely related to the masses of the PBHs formed through critical collapse. Indeed, we can see that, while PTA experiments set constraints in the (sub)-solar mass range for PBHs~\cite{Iovino:2024tyg}, the asteroidal mass range, i.e., $m_{\rm PBH} \in [10^{-17},10^{-5}]$ $\msun$, can be tested by space-based interferometers such as LISA and TIANQin.

Our results show that the SIGW signal can, in principle, bound the amplitude of the primordial spectrum up to the order $\sim 10^{-5}$ at the SKA and LISA scales for a peaked power spectrum and up to the order $\sim 10^{-4}$ in the broader case. This behaviour is reflected for all the detectors considered, even if for AEDGE we find that the prospects are between orders $10^{-5}$ and $10^{-6}$. Interestingly, the CE and ET bounds are of the order $\sim 10^{-4}$ for a peaked spectrum. The difference between the broad and peaked spectra can be understood by recalling that for narrow spectra around the peak we have $\Omega_{\rm GW}^{\rm Gauss}(f_{\rm peak}) \simeq 2.6$, while for broad spectra $\Omega_{\rm GW}^{\rm Gauss}(f_{\rm peak}) \simeq 0.125/\Delta^2$~\cite{Pi:2020otn}. Thus, the wider the scalar power spectrum, the lower the amplitude of the GW signal, and this originates from a dilution of the modes due to the broadness of the spectrum. Thus, a broad spectrum would imply a lower amplitude of the integral in the Gaussian case, and as a consequence, the amplitude of fluctuations $A$ must be higher to match the SNR of each detector. The behaviour is different for spectral distortions since those bounds are obtained by integrating the scalar power spectrum over the whole $k$ range. Thus, the largest variations can be seen in the tails. Our prospects thus suggest that, even in the case of a non-detection of a SIGW background, future GW interferometers can constrain the scalar power spectrum across 10 decades in $k$.

\subsection{The LISA case}

\begin{figure}
    \centering
    \includegraphics[width=1\linewidth]{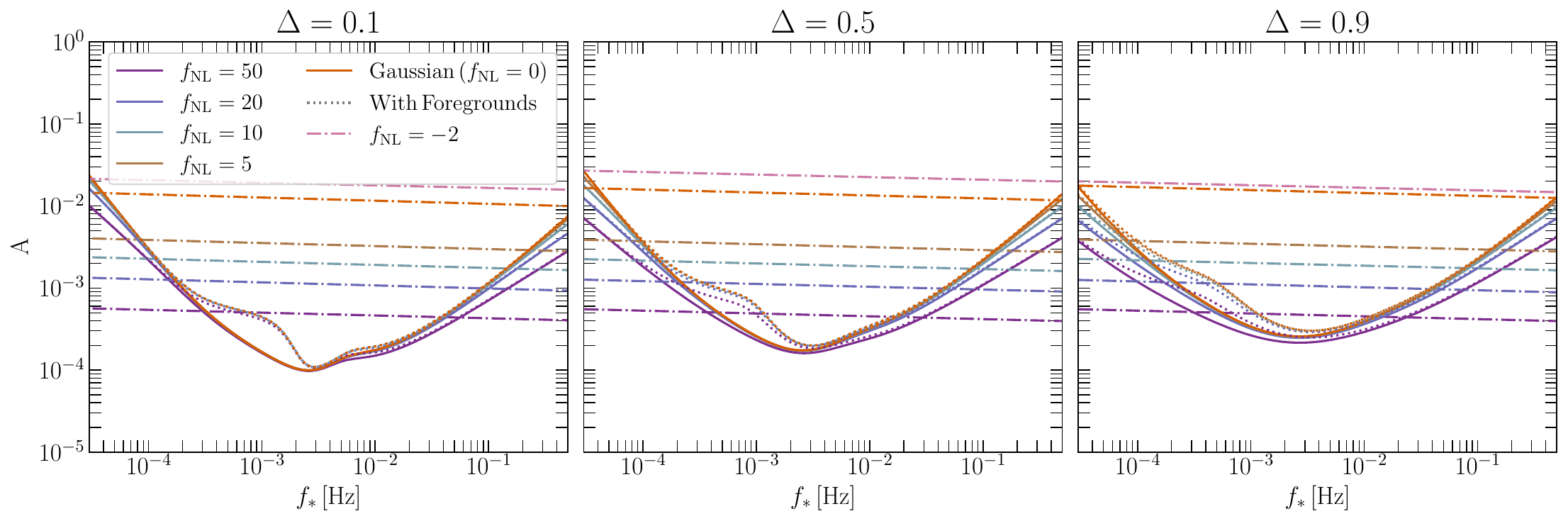}
    \caption{Projected sensitivities on the amplitude of the scalar power spectrum a detection of a SIGW signal from LISA, with ${\rm SNR} = 10$ and 4 years of observation time. Solid lines represent the bounds when the presence of foregrounds is neglected, while dotted lines represent the bounds when foregrounds are included. The dot-dashed lines indicate $f_{\rm PBH} = 1$ for different $f_{\rm NL}$.}
    \label{fig:LISA_NG}
\end{figure}

To accurately estimate the prospects for LISA to probe PBHs via SIGW, it is crucial to consider the effect of NG. This is illustrated in Fig.~\ref{fig:LISA_NG}, where we show the amplitude $A$ required to reach SNR=10 for a given $k_*$, width $\Delta$ and $f_{\rm NL}$ together with the amplitude $A$ corresponding to $f_{\rm PBH} = 1$. The sensitivity curves were obtained by computing the SIGW spectrum at the leading non-Gaussian order, where it can be decomposed as 
\bea
\label{eq::GW+NG}
    h^2 \Omega_{\rm GW}(f,A) = 
    A^2 h^2\Omega_{\rm GW,0}^{\rm Gauss}(f) + A^3 f_{\rm NL}^2 h^2\Omega^{\rm NG}_{\rm GW,0}(f)\,
\eea
and then determining $A$ from Eq.~\eqref{Eq:SNR}. We assumed 4 years of observation time. The solid lines show prospects assuming only the LISA instrumental noise, while the dashed lines also account for the astrophysical foregrounds. The colours indicate $f_{\rm NL}\in[-2,50]$ with the Gaussian case $f_{\rm NL} = 0$ shown in black. 

Fig.~\ref{fig:LISA_NG} shows that NG corrections have an almost negligible impact on the sensitivity curves. The only appreciable difference can be seen in the tails, since the NG contribution is more pronounced at larger amplitudes. This can be quantitatively understood as the relative contribution of NG becomes relevant roughly when $(3/5 f_{\rm NL})^2A \gtrsim 1$. For example, when $f_{\rm NL} = 50$, we expect to see an effect when $A>10^{-3}$, which is in decent agreement with Fig.~\ref{fig:LISA_NG}. Nevertheless, we observe that this estimate varies with the spectral shape. In particular, NGs can have a stronger impact for wider spectra. For instance, $\Delta = 0.9$, there is a visible increase in sensitivity at all frequencies when moving from the Gaussian case to the $f_{\rm NL} = 50$ case. When $\Delta = 0.9$ such an effect is seen only in the tails

The shape of the sensitivity curves can be understood by recalling that the LISA sensitivity curve has a minimum around $f\sim 10^{-3}$ Hz, as well as the fact that for spectra peaked on the left of the minimum, LISA is sensitive only to a small UV part of the spectrum, while for spectra peaked on the right, LISA is sensitive to the whole IR tail. 
Thus, with an observation time of 4 years, LISA is expected to reach amplitudes as low as $A \sim 10^{-4}$ around $f\sim 10^{-3}$ Hz. For such low amplitudes, the effect of NG is effectively negligible, as explained above, and there is a mild weakening of the sensitivity to wider power spectra since the peak of the SIGW spectrum is smoother in that case.
In the case of negative NGs, while the SIGW is insensitive to the sign of $f_{\rm NL}$, the amplitude corresponding to $f_{\rm PBH}=1$ is larger compared to the Gaussian case. 

The sensitivity to $A$ is further reduced by astrophysical foregrounds. These are generated by the superposition of numerous weak and unresolvable signals from various astrophysical sources (see, e.g.~\cite{Schneider:2000sg, Farmer:2003pa, Regimbau:2011rp,LISA:2022yao,Pozzoli:2023kxy,Babak:2023lro,Staelens:2023xjn,Toubiana:2024qxc}). Two guaranteed components are expected to be present in the LISA band: an extra-galactic GW foreground sourced by neutron star (NS) and BH binaries~\cite{LIGOScientific:2019vic}, and a galactic foreground, generated by white dwarf binaries~\cite{Evans:1987qa,Bender:1997hs}. We account for the foregrounds by adding them to $\Omega_{\rm noise}(f)$ in Eq. \eqref{Eq:SNR}, adapting their shape from Ref.~\cite{LISACosmologyWorkingGroup:2025vdz}. However, we note that the spectral shape and amplitude foregrounds are subject to sizeable uncertainties that stem from our limited knowledge of the formation and distribution of its sources (for more details, see Ref.~\cite{LISACosmologyWorkingGroup:2025vdz} and~\cite{Caprini:2024hue,Blanco-Pillado:2024aca,LISACosmologyWorkingGroup:2024hsc,Kume:2024xvh}).

Finally, consider the amplitudes $A$ corresponding to $f_{\rm PBH} = 1$ shown by the dot-dashed lines in Fig.~\ref{fig:LISA_NG}. All scenarios above these lines are excluded by PBH overproduction. Crucially, we can observe how, unlike SIGW, NG significantly impacts the abundance of PBH. For example, the required amplitude $A$ for PBH overproduction is reduced by almost an order of magnitude when moving from the Gaussian case $f_{\rm NL} = 0$ to $f_{\rm NL} = 50$. This effect is especially notable for the wider spectra $\Delta = 0.9$, in which case LISA's ability to prove viable PBH scenarios ($f_{\rm PBH} < 1$) is drastically reduced from $f_{\rm NL} = 0$ to $f_{\rm NL} = 50$. Moreover, as seen in Fig.~\ref{fig:Abu}, the suppression of PBH production due to negative NGs is strongest when $f_{\rm NL} \approx -2$, and thus the required amplitude to reach $f_{\rm PBH}=1$ begins to decrease again when $f_{\rm NL} \lesssim -2$. We added the corresponding line in the Figure as a reference. We will return to the implications for PBHs in Sec.~\ref{sec:analysis}. 

\subsection{Prospects for SIGW detection}
\begin{figure}[t!]
    \centering
    \includegraphics[width=0.65\textwidth]{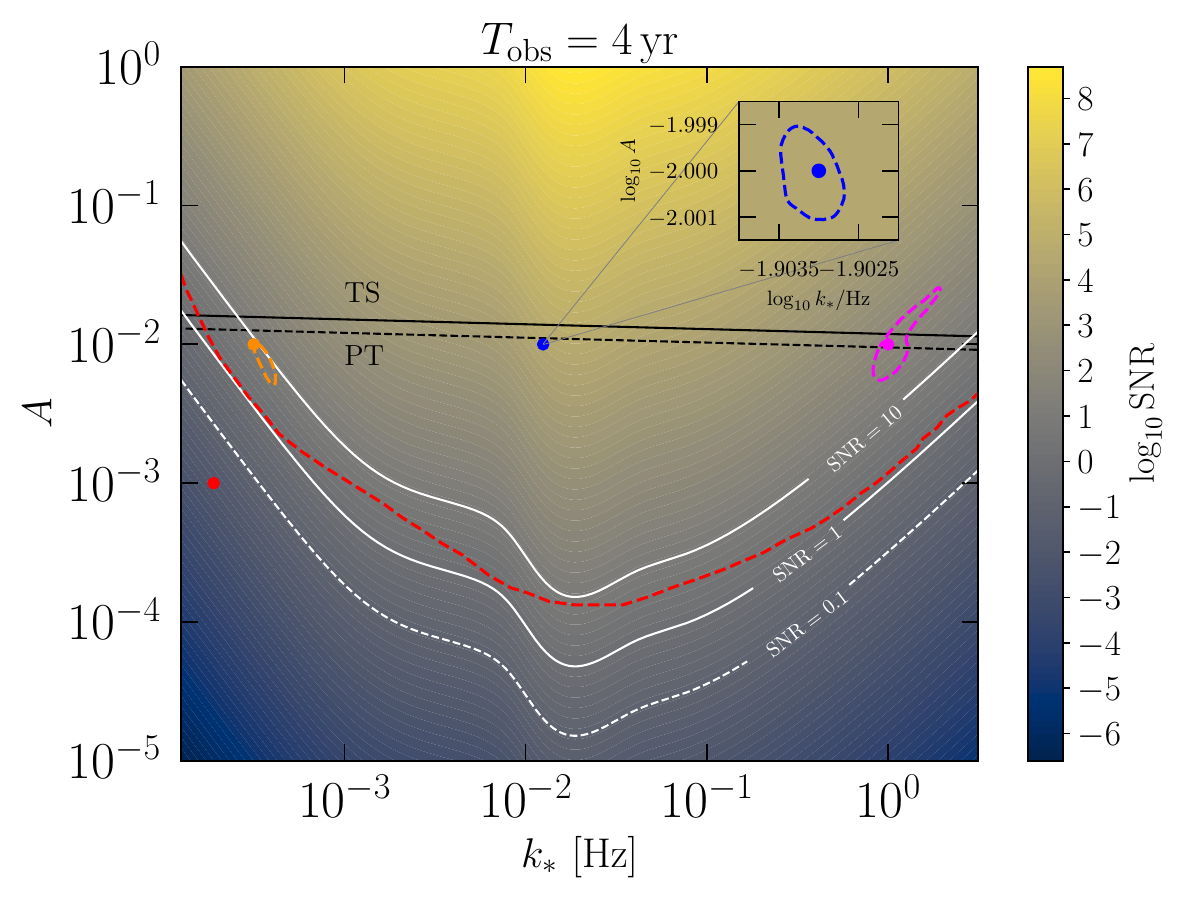}
    \caption{Contour plot of the SNR computed in the Gaussian approximation ($f_{\rm NL} = 0$), assuming a log-normal power spectrum with width $\Delta = 0.3$, shown as a function of the peak position $k_*$ and the amplitude $A$. The panel also includes four selected configurations (colored dots) used as injected signal for the MCMC analysis, along with their corresponding 2D 95\% confidence contours (dashed colored lines). Two black horizontal lines indicate the threshold values of $A$, as a function of $k_*$, above which PBH overproduction occurs, according to the TS (solid) and PT (dashed) formalisms.}
    \label{Fig:SNR_k*_A}
\end{figure}
In Fig.~\ref{Fig:SNR_k*_A}, we estimate the detectability of a SIGW assuming an observation time of 4 years 
and show the SNR, including astrophysical foregrounds, for the SIGW at each $k_*$ and $A$ for a fixed width $\Delta = 0.3$. As expected, for a fixed $k_*$, a higher amplitude $A$ implies a higher SNR. The behaviour for varying $k_*$, as well as the constant SNR lines, can be understood from the shape of the LISA sensitivity curves and the shape of the spectrum. For completeness, we also show the $f_{\rm PBH} = 1$ lines in the TS and PT formalism.

\begin{table}[h]
\centering
\small
\begin{tabular}{l c c | c c c c c}
\toprule
       & $A$       & $k_*/\rm Hz$     & $\log_{10} A$       & $\log_{10}k_*/\rm Hz$   & $\log_{10}\Delta$ & $|f_{\rm NL}|$ & {\rm SNR}\\
\midrule
\textcolor{orange}{B.I}   & $10^{-2}$ & $10^{-3.5}$   &  ${-2.15\pm0.05}$ &  ${-3.43\pm0.03}$ &${-0.7\pm0.1}$ & $<6.45$ & 8\\
\textcolor{blue}{B.II}  & $10^{-2}$ & $10^{-1.903}$ &     ${-2\pm0.0004}$                    & ${-1.9031\pm0.0002}$   & ${-0.5231\pm 0.0003}$   &  $<0.8$&   $2.5\cdot10^4 $          \\
\textcolor{magenta}{B.III} & $10^{-2}$ & $1$    &               ${-2.0\pm0.2}$          &   ${0.06\pm0.03}$  & ${-0.4\pm0.2}$ & $<3$ & $10^2$             \\
\textcolor{red}{B.IV}  & $10^{-3}$ & $10^{-3.7}$  &      $<-3.08$                   & $-$ & $-$ &$<20$ &$1.6\cdot10^{-2}$
\\
\bottomrule
\end{tabular}
\caption{Benchmark cases used to realise Fig.~\ref{Fig:SNR_k*_A}. We fixed $\Delta = 0.3$ and $f_{\rm NL} = 0$. The injected values are shown in the first 2 columns and the remaining columns show the 68\% confidence intervals obtained from the MCMC and the SNR.}
\label{Tab::benchmark}
\end{table}
The coloured dots in Fig.~\ref{Fig:SNR_k*_A} show three benchmark cases described in Table~\ref{Tab::benchmark} with the corresponding 2D 95\% confidence contour in the $A-k_*$ plane\footnote{In the table, we do not report the MCMC results for $k_*$ and $\Delta$ because, given the low signal amplitude, the posterior distribution of $\Delta$ is essentially flat over the entire prior range, while the posterior of $k_*$ exhibits peaks on both sides of the LISA central frequency, corresponding to regions where the detector sensitivity is reduced.}. These confidence regions have been obtained by running several MCMC scans using \texttt{SIGWAY}~\cite{LISACosmologyWorkingGroup:2025vdz}.
The cases B.I and B.III show the behaviour of the inference when the signal lies near the edges of the LISA sensitivity curve, with an SNR of approximately $\mathcal{O}(10)$. The specific shape of the contours depends both on the position of the signal in the parameter space and on the particular noise realisation used in the mock analysis.

B.II, in contrast, was chosen to lie near the central part of the LISA sensitivity curve, which represents the optimal region for detecting a SIGW signal. As shown in Fig.~\ref{Fig:SNR_k*_A}, the SNR in this case is of the order of $\sim 10^4$: the resulting constraints are significantly tighter, yielding a relative error at the per-mille level.

Finally, B.IV lives in an extremely low-SNR region: by Fig.~\ref{Fig:SNR_k*_A}, its SNR is $\mathcal{O}(10^{-2})$. This is reflected in the behaviour of the 95\% level contours, which are non-constraining but infer only an upper bound on $A$ across the whole LISA frequency range. The U-shaped contour stays between the ${\rm SNR}=1$ and ${\rm SNR}=10$ lines, while following the ${\rm SNR}=1$ curve at the edges and being closer to the ${\rm SNR}=10$ curve in the centre. Understandably, its shape resembles the LISA sensitivity curve, because when injecting an unobservable signal, its $95\%$ confidence region should exclude points where a signal would be detected. 

While an undetectable signal might not be the most interesting, the resulting confidence region provides a useful comparison between the sensitivity obtained from SNR and MCMC analyses and shows that both methods give comparable sensitivities. However, we must remark that a complete MCMC estimate for the sensitivity in the $(k_*,A)$ plane would instead entail a computationally expensive scan, detailing at each point whether an injected signal would be detected at the required confidence level.

\section{PBH prospects for LISA}
\label{sec:analysis}

\subsection{Sensitivity to PBH abundance}

\begin{figure}
    \centering
    \includegraphics[width=0.99\linewidth]{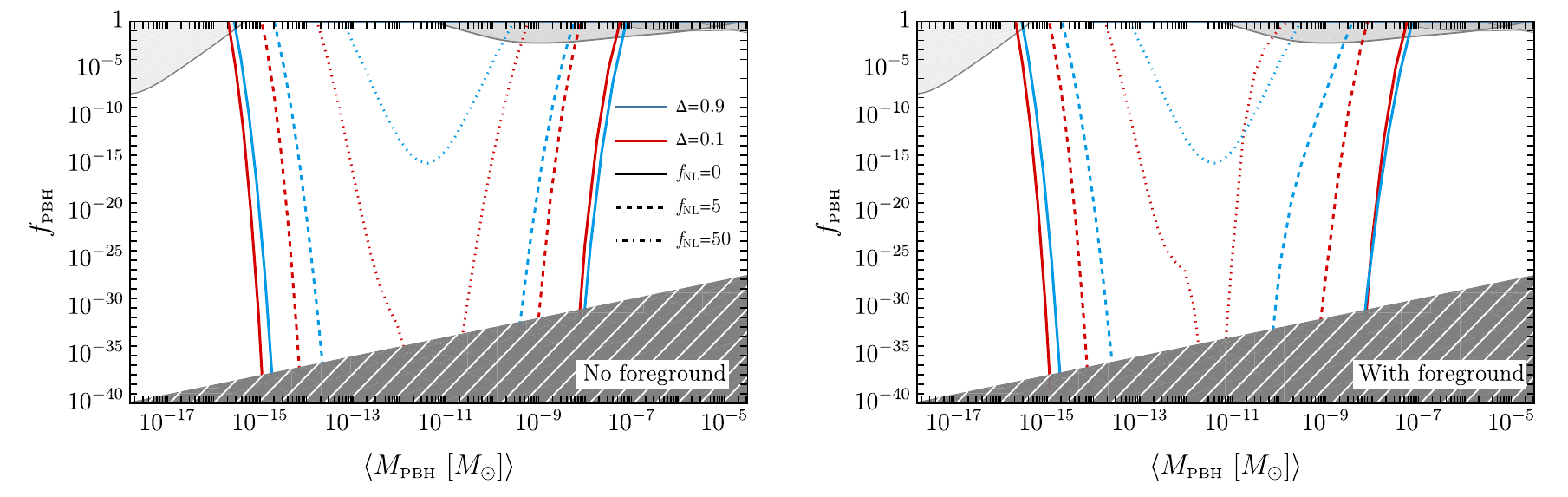}
    \caption{ The LISA projected constraints on the PBH abundance using TS approach, assuming several benchmark cases for the log-normal power spectrum width and primordial NGs for the entire observation time of the experiment and SNR=10, without (left panel) and with (right panel) an astrophysical foreground.
}
    \label{fig:ConsFpbh}
\end{figure}

We consider the projected constraints on the SIGW signal from future LISA observations. The influence of these constraints on the shape of the primordial power spectrum and the required level of primordial NG is shown in Fig.~\ref{fig:ConsFpbh}. 

We begin by highlighting that the evaporation constraints for extended mass functions differ significantly from those in the monochromatic case, due to the finite width of the mass distribution~\cite{Carr:2017jsz}. While monochromatic scenarios typically impose evaporation bounds up to $\langle M_{\rm PBH} \rangle \simeq 10^{-16}$ $\msun$ (see also~\cite{Saha:2021pqf,Laha:2019ssq,Ray:2021mxu,Khan:2025kag}; and constraints from EDGES~\cite{Mittal:2021egv}, CMB~\cite{Clark:2016nst}, INTEGRAL~\cite{Laha:2020ivk,Berteaud:2022tws}, the 511 keV line~\cite{DeRocco:2019fjq,Dasgupta:2019cae}, Voyager~\cite{Boudaud:2018hqb},EGRB~\cite{Carr:2009jm} and lyman$-\alpha$~\cite{Saha:2024ies}), the presence of a critical tail in extended mass functions pushes these limits further, up to $\langle M_{\rm PBH} \rangle \simeq 4 \times 10^{-16}\msun$ as can be seen in Fig.~\ref{fig:ConsFpbh}. This shift significantly affects the minimum level of primordial NGs required to preserve any viable parameter space in which the entirety of dark matter could consist of PBHs. The constraints for heavier PBHs, which are dominated by microlensing, including from Hyper Suprime-Cam~\cite{Niikura:2017zjd} and OGLE~\cite{Niikura:2019kqi,Mroz:2024wag,Mroz:2024wia}, do exclude slightly lower masses when the width of the mass function has been accounted for, although the effect is milder. 

The shaded region in Fig.~\ref{fig:ConsFpbh} shows scenarios in which less than a single PBH would exist within the current Hubble volume. This bound, dubbed the cosmological incredulity limit, is given by~\cite{Carr:1997cn, Carr:2020erq}
\be
    f_{\rm PBH} \cdot 4\pi \Omega_{\rm DM} M_{\rm pl}^2/H_0
    > \langle M_{\rm PBH} \rangle\,.
\ee
Below this limit, PBHs become observationally irrelevant.

As is evident from Fig.~\ref{fig:ConsFpbh}, the prospective sensitivity to SIGW implies nearly vertical sensitivity contours in the PBH parameter space. In the absence of significant primordial NG, any conceivable amount of PBHs in the asteroid-mass window would imply a detection of a SIGW by LISA. A non-detection, on the other hand, would exclude the existence of even a single PBH within the Hubble volume. 

Such prospective constraints can be relaxed when $f_{\rm NL}$ sufficiently large, allowing PBHs up to masses $\sim 10^{-9} M_\odot$ ($\sim 10^{-10} M_\odot$) with $f_{\rm NL}=5$ (50) to remain viable in the case of non-detection of a SIGW by LISA. In absence of NGs, the constraints are limited at masses $\sim 10^{-7} M_\odot$. A mass range that could be probed by future interferometers such as the Einstein Telescope (ET)~\cite{Pujolas:2021yaw,Branchesi:2023mws,Franciolini:2023opt}.

This behaviour can be understood from Fig.~\ref{fig:LISA_NG}. In the absence of astrophysical foregrounds, the projected upper bounds on the scalar power spectrum amplitude are nearly degenerate for all $f_{\rm NL}$ considered, except in the low- and high-frequency tails, where deviations become noticeable for large NGs ($f_{\rm NL} \gtrsim 50$). Conversely, the amplitude required to reach $f_{\rm PBH} \simeq 1$ decreases monotonically with increasing $f_{\rm NL}$. This results in a narrowing of the allowed parameter space at large $f_{\rm NL}$, effectively reopening the PBH window in the asteroid-mass regime only when NG is large enough. Additionally, while SIGWs are sensitive to $|f_{\rm NL}|$ due to their quadratic dependence, PBH formation exhibits a strong dependence on the sign of $f_{\rm NL}$. Consequently, in Fig.~\ref{fig:ConsFpbh} we restrict our analysis to the positive $f_{\rm NL}$ cases, as negative NG only leads to more stringent bounds.

In summary, the LISA sensitivity to PBH abundance is only weakly dependent on the width $\Delta$ of the power spectrum, with slightly more stringent constraints for narrow spectra ($\Delta = 0.1$). As $f_{\rm NL}$ increases, the exclusion region tightens, reflecting the importance of NGs in the computation of the constraints.

This picture changes partially when astrophysical foregrounds are included. Indeed, while in the Gaussian or weakly NG cases the situation is barely untouched,  in the presence of large NGs, the effect of foregrounds becomes more pronounced, with the constraints relaxing considerably depending on $\Delta$. This is because, as shown in Fig.\,\ref{fig:LISA_NG}, the projected bounds on the power spectrum amplitude in the presence of foregrounds with large NGs differ significantly between narrow and broad cases.

\subsection{Inference on the PBH abundance from a SIGW signal}
\begin{figure}
    \centering
    \includegraphics[width=0.7\textwidth]{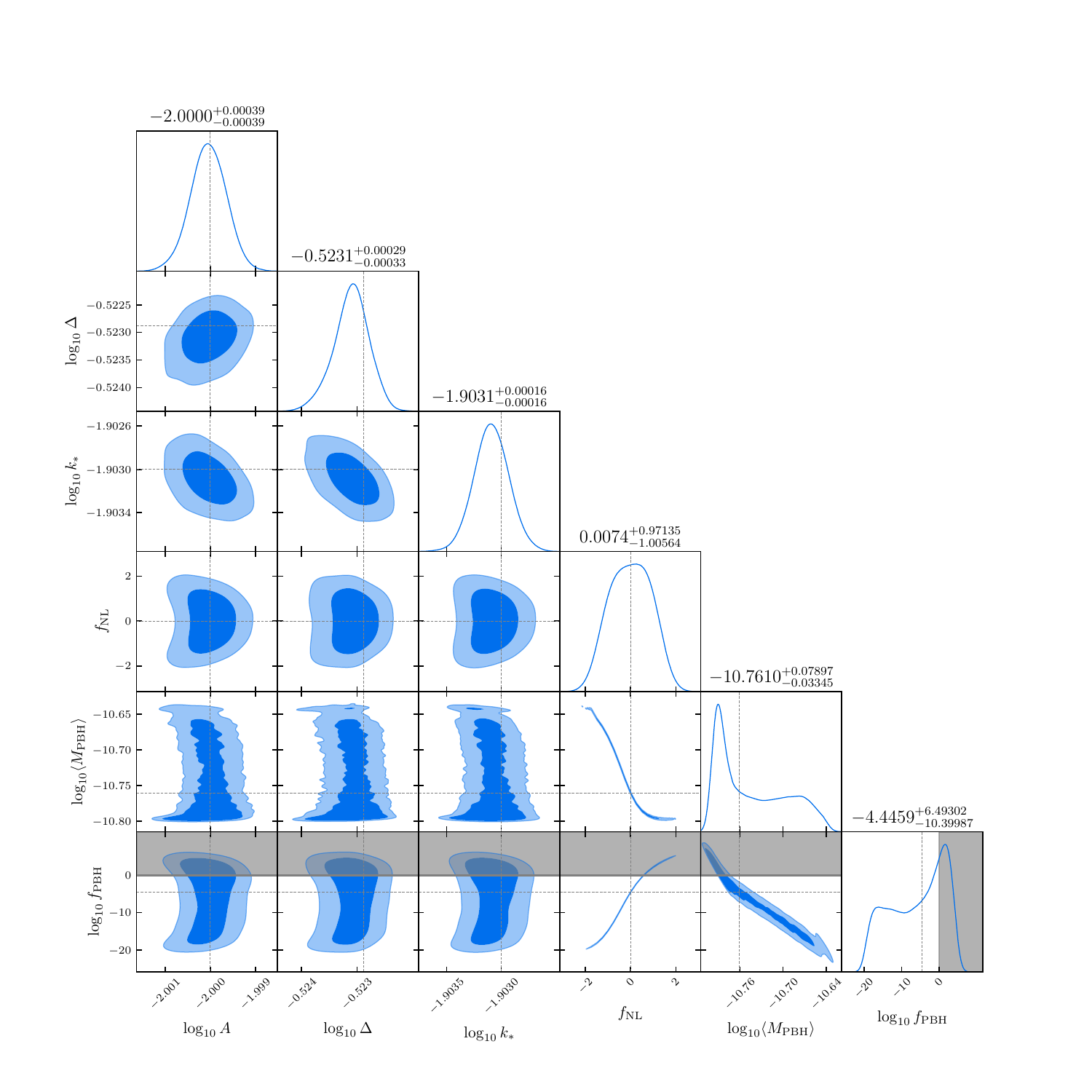}
    \caption{Posterior distributions for an injected signal without any primordial NGs, $ f_{\mathrm{NL}}=0$, and assuming a log-normal power spectrum with $A=10^{-2}$, $\Delta=0.3$ and $k_*=1.25\cdot10^{-2}$ $\rm{Hz}$.}
    \label{fig:Inference}
\end{figure}

As outlined in Sections~\ref{sec:SIGWs} and~\ref{sec:PBHs}, SIGW and PBH production is highly sensitive to the detailed shape and structure of the primordial power spectrum. To estimate the uncertainty in the prediction of the abundance and mean mass of PBH, we proceed as follows. We first evaluate the capability of LISA to constrain the amplitude $A$, the width $\Delta$, the central frequency $f_*$ (or equivalently $k_* = 2\pi f_*$), and $f_{\rm NL}$. For the purpose, we use the \texttt{SIGWAY}~\cite{LISACosmologyWorkingGroup:2025vdz} and the \texttt{SGWBinner} pipeline \cite{Caprini:2019pxz, Flauger:2020qyi} pipeline to generate mock data by injecting a SIGW signal and then performing Bayesian inference.

The posteriors for the derived quantities $f_{\rm PBH}$ and $\langle M_{\rm PBH} \rangle$ can be inferred from the posteriors of $\{A, \Delta, f_*, f_{\rm NL}\}$. Indeed, computing them directly within the MCMC analysis would be computationally prohibitive due to the presence of nested integrals at each MCMC step. Instead, we construct a multi-dimensional grid of $f_{\rm PBH}$ and $\langle M_{\rm PBH} \rangle$ computed for a range of $A$, $\Delta$, $k_*$, and $f_{\rm NL}$, use it to interpolate the MCMC outcomes and extract the corresponding PBH-related quantities. The grey region in Fig.~\ref{fig:Inference}. indicates PBH overproduction.

As benchmark points, we inject a scenario with $A = 10^{-2}$, $\Delta = 0.3$, $k_* = 2\pi f_* = 1.25 \cdot 10^{-2}$ Hz (corresponding to $f_* = 2 \cdot 10^{-3}$ Hz) and $f_{\rm NL} = 0$, which yields a well resolved SIGW spectrum within the LISA frequency band. The results of the MCMC analysis are shown in Fig.~\ref{fig:Inference} and in the second row of Table~\ref{Tab::benchmark}. 
The parameters of the power spectrum are constrained with high precision, with relative errors at the level of $\sim 10^{-3}$. The marginalised posterior of $f_{\rm NL}$ is centred around zero and exhibits symmetry, as expected due to the quadratic dependence of the induced GW spectrum on $f_{\rm NL}$. This symmetry is also evident in the shape of the 2D posteriors involving $f_{\rm NL}$. Our results are consistent with those of~\cite{LISACosmologyWorkingGroup:2025vdz} and with previous forecasts in~\cite{Perna:2024ehx}.

The posterior of the mean PBH mass $\langle M_{\rm PBH}\rangle$ is confined within a relatively narrow range. On the other hand, the posterior of the PBH abundance $f_{\rm PBH}$ spans nearly 30 orders of magnitude. This behaviour can be understood as follows: the mean mass is only mildly sensitive to NG but depends more strongly on the characteristic frequency and width of the power spectrum. In contrast, the PBH abundance is highly sensitive to $f_{\rm NL}$, which can significantly enhance or suppress it. Moreover, as shown in~\cite{Iovino:2024tyg,Franciolini:2023pbf}, the impact of positive and negative $f_{\rm NL}$ on the abundance is asymmetric. Thus, even if the posterior on $f_{\rm NL}$ appears to be quite narrow, the associated uncertainty is sufficient to induce large variations in the predicted PBH abundance. We have verified that reducing the uncertainty in $f_{\rm NL}$ would significantly improve the reconstruction of both PBH-related quantities.

We further highlight that the plots showing the marginalised 2D posterior of $f_{\rm PBH}$ (and $\langle M_{\rm PBH}\rangle$) with the power spectrum parameters appear almost identical. This occurs because their specific correlations are entirely masked by the uncertainty in $f_{\rm NL}$. Thus, treating $f_{\rm NL}$ as fixed is unjustified and can severely underestimate the uncertainty in $f_{\rm PBH}$.

\subsection{Uncertainty in primordial non-Gaussianity}

We investigate to what extent LISA is capable of constraining primordial NG for different values of amplitude $A$ and peak frequency $f_*$. We explore the entire frequency-band probed by LISA for different values of the amplitude, similarly to Fig.~\ref{Fig:SNR_k*_A}. 

\begin{figure}[t!]
    \centering
    \includegraphics[width=0.7\textwidth]{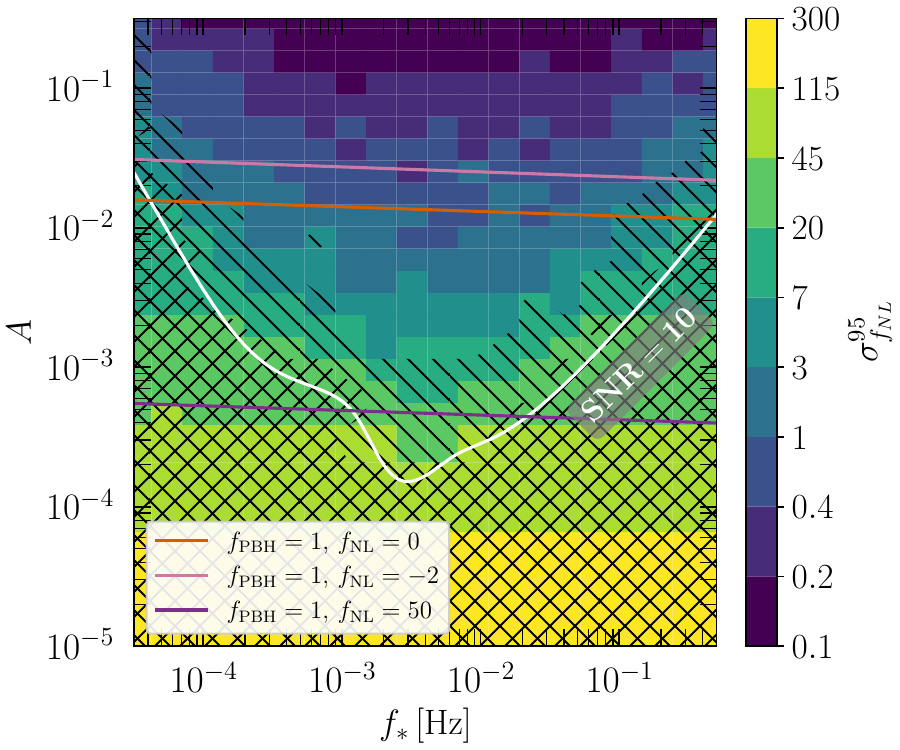}
    \caption{Contour plot showing the 95\% confidence level on $f_{\rm NL}$ when a Gaussian signal is injected. We also report the lines corresponding to $f_{\rm PBH}=1$ for different values of $f_{\rm NL}$.}
    \label{Fig:fNL_k*_A}
\end{figure}

We sample the posterior distribution using a Bayesian Monte-Carlo method for each $(A,f_*)$ on a grid. In all cases, we inject a Gaussian signal; however, in inference, we do not make any assumptions about the specific value of $f_{\rm NL}$, thus leaving it as a free parameter. We adopt a flat prior in the range $f_{\rm NL}\in{1}/{\sqrt{A}}[-1,1]$, wide enough to ensure that all values satisfying the perturbativity condition are included. For each run, we compute the 95\% percentile of the distribution that we report in Fig.~\ref{Fig:fNL_k*_A}. We also display the line corresponding to $\rm SNR=10$ as a reference for the detectability of the signal.

The region where the 95\% constraints are weaker than the perturbativity condition~\eqref{eq:PerturbCri1}, that is, when $(f_{\rm NL}^{95\%})^2A>0.1$ are indicated by hatched lines slanting to the left. In that case, a small part of the posterior contains a region where the theoretical modelling of the SIGW spectrum becomes unreliable. The cross-hatched region shows where $f_{\rm NL}>0.90/\sqrt{A}$. This corresponds to a posterior that is almost as flat as the prior (or matches the prior, as happens, for example, in the lowest part of the plot). More importantly, this region encloses those values where most of the posterior reaches large enough $f_{\rm NL}$ to violate perturbativity, and thus the analysis ceases to be fully reliable. Additionally, this region is always below the detectability line, where it is expected that LISA's ability to constrain $f_{\rm NL}$ is negligible as well. 

The map we provide allows one to pinpoint the expected uncertainty on $f_{\rm NL}$ across the entire LISA range. This is crucial for assessing LISA’s sensitivity to primordial NG, as well as for identifying the limitations that arise from our incomplete understanding of how to model non-Gaussian effects in regimes where perturbativity breaks down.

Interestingly, the same map is of practical use for understanding how the uncertainty of $f_{\rm NL}$ might affect the uncertainty on $f_{\rm PBH}$. First, for the purpose of the following analysis, we highlight that all points above the detectability line lie in the range $\sigma^{\rm 95\%}\lesssim 50$, which agrees with the largest $f_{\rm NL}$ depicted in Fig.~\ref{fig:ConsFpbh}. We also show the  $f_{\rm PBH}=1$ evaluated using TS,curves for different benchmarks: the Gaussian case with $f_{\rm NL}=0$, as well as $f_{\rm NL}=50$ and $f_{\rm NL}=-2$. Since $f_{\rm NL}=-2$ yields the strongest suppression of PBH formation, it acts as a limiting case: any value of $A$ above this line would overproduce PBHs and is therefore strictly excluded.

To illustrate the usage of the table, let us now consider the injected values in Fig.~\ref{fig:Inference} and $(A,f_*)=(10^{-2},2\cdot 10^{-3})$ as an example. The forecast uncertainty shown in the graph at the confidence level of 95\% lies in the range of 1–3 (the exact value is $\sim 1.70$). Consequently, even if the injected value is $f_{\rm NL} = 0$, the uncertainty is wide enough to allow values up to almost $\sim \pm 2$. The line corresponding to $f_{\rm PBH}=1$ for $f_{\rm NL}=0$ lies above the point considered, indicating that a detection of those value with $f_{\rm NL}=0$ would constrain the PBH abundance. However, the confidence level is wide enough to allow $f_{\rm NL}=-2$. The corresponding line for $f_{\rm PBH}=1$ lies above the previous one, thus suggesting that the constraints on the abundance would be even stronger. However, the line corresponding to $f_{\rm NL}=2$ is also allowed. The position of this line, as hinted from Fig.~\ref{Fig:SNR_k*_A}, would be below the point, implying an overproduction of PBHs. This behaviour was shown explicitly in Fig.~\ref{fig:Inference}, where the uncertainty in $f_{\rm NL}$ translates into an uncertainty in $f_{\rm PBH}$ that spans almost 30 orders of magnitude! It is important to stress here that, if a signal is detected, nearly all of this variation in  $f_{\rm PBH}$ would stem from the uncertainty in $f_{\rm NL}$. In the end, despite being able to reconstruct the parameters with relatively small errors, the uncertainty in $f_{\rm NL}$ completely compromises any conclusions regarding the abundance of PBHs.

Let us now conclude by considering another case, with the same $f_*$, but $A=5\cdot10^{-4}$. The point would be significantly below the line that saturates the PBH abundance for $f_{\rm NL}=0$, implying strong constraints that would exclude a large range of masses in the PBH abundance. However, the uncertainty permits $f_{\rm NL}$ in the range $20-45$. In this case, the corresponding lines for $f_{\rm PBH}=1$ would always lie above the considered point (see also Fig.~\ref{Fig:SNR_k*_A}). This behaviour can also be seen in Fig.~\ref{fig:ConsFpbh}: when $f_{\rm NL} =0$ the window is completely closed, but as it is increased, some ranges of masses could still allow $f_{\rm PBH}=1$. Despite the uncertainty in $f_{\rm NL}$, we would still be able to constrain the abundance, but the strength of the constraints is limited by the uncertainty on primordial non-Gaussianity. In addition, we note that the point lies in a region where part of the distribution falls outside the perturbative regime, so not only might $\sigma^{\rm 95\%}$ differ a bit, but also the estimate of the PBH abundance could receive notable corrections.

\section{Conclusions}
\label{sec:Conc}

In this study, we have explored the capability of LISA to constrain the existence and abundance of asteroid-mass PBHs as a dark matter candidate through the associated SIGW background. Exploiting the fact that primordial scalar perturbations generate both PBHs and SIGWs, we derived the consequences that a mHz-band detection or non-detection of SIGWs would have for the PBH abundance. Whereas SIGWs depend quadratically on $f_{\mathrm{NL}}$ (i.e., only on $|f_{\mathrm{NL}}|$), PBH formation responds exponentially to both the sign and size of the non-Gaussianity, since non-Gaussianity strongly modifies the tails of the distribution of curvature perturbations.

We derived LISA’s sensitivity to the amplitude $A$ of the scalar power spectrum for different choices of its width $\Delta$ and the degree of non-Gaussianity. For purely Gaussian perturbations, LISA is can be sensitive to $A \gtrsim 10^{-4}$ for narrow spectra, assuming four years of observation, regardless of astrophysical foregrounds, which would moderately weaken the bounds. 

LISA's sensitivity to the power spectrum translates into prospective constraints on the PBH abundance. In the Gaussian case, a non-detection of SIGWs in the LISA band would essentially exclude any conceivable amount of PBHs in the asteroid-mass range ($10^{-16} M_\odot \lesssim M_{\mathrm{PBH}} \lesssim 10^{-7} M_\odot$). With sizable positive non-Gaussianity ( $f_{\mathrm{NL}} \gtrsim 5$), the limits loosen substantially, leaving room for PBHs to comprise all of dark matter in the low mass end of the asteroid mass window. Astrophysical foregrounds can also soften the prospective constraints, but their effect is much milder when compared to the effect of non-Gaussianity.

To characterize the SIGW–PBH connection more precisely, we performed a Bayesian parameter estimation using mock SIGW signals in the LISA band. We find that in the middle of the frequency band LISA can pin down the curvature power spectrum parameters (amplitude, width, peak frequency) to subpercent accuracy. Although the mean PBH mass is also recovered  reliably, the inferred PBH abundance $f_{\mathrm{PBH}}$ remains poorly determined because it is extremely sensitive to even small changes in $f_{\mathrm{NL}}$. For instance, we show that uncertainties in $f_{\mathrm{PBH}}$ can vary by 30 orders of magnitude for otherwise well reconstructed signals (SNR = $\mathcal{O}(100)$) due to the uncertainties in $f_{\mathrm{NL}}$.

Because non-Gaussianity-related uncertainties are the dominant obstacle to drawing robust conclusions about PBHs, we examined how well LISA can test primordial non-Gaussianity over its full frequency range. We find that, although LISA could strongly bound PBH abundance in certain parameter regions, the uncertainty in $f_{\rm NL}$ severely hampers such inferences for almost all SIGW scenarios accessible to LISA. We therefore construct a map that quantifies these uncertainties and delineates where perturbative calculations break down and theoretical predictions lose reliability. In summary, even with reasonably accurate parameter reconstruction, the limited knowledge of $f_{\rm NL}$ can substantially weaken LISA's ability to measure $f_{\rm PBH}$.

Overall, our results highlight the critical importance of including primordial non-Gaussianity and, to a smaller extent, astrophysical foregrounds when interpreting future SIGW data. While LISA alone may not be sufficient to definitively confirm or rule out the PBH dark matter scenario without additional constraints on non-Gaussianity, it will still significantly narrow the viable parameter space. A non-detection of SIGWs would either exclude large regions of the model space or point to the necessity of substantial non-Gaussianity in the early Universe. Conversely, a detection would provide compelling evidence of PBH formation processes tied to early-universe physics.

\acknowledgments
\noindent
We thank A. Ricciardone, A. Riotto, and V. Vaskonen for useful discussions.
H.V. and G.P. are supported by the Estonian Research Council grants PSG869, RVTT7, KOHTO34, TARISTU24-TK3, TARISTU24-TK10 and the Center of Excellence program TK202. G.P. acknowledges partial financial support by ASI Grant No. 2016-24-H.0 and thanks Fondazione Angelo Della Riccia and Fondazione Aldo Gini for financial support.

\bibliographystyle{JHEP}
\bibliography{refs}

\end{document}